\documentclass[journal,10pt]{IEEEtran}
\usepackage{lineno}
\usepackage{cite}
\usepackage{hyperref}
\usepackage{amsmath,amssymb,amsfonts}
\usepackage{amsthm}
\usepackage{mdwmath}
\usepackage{mdwtab}
\DeclareMathOperator*{\argmax}{argmax}

\usepackage{graphicx}
\usepackage{textcomp}
\usepackage{xcolor}
\usepackage{graphicx}
\usepackage{float}
\usepackage{subfigure}
\usepackage{mathrsfs}
\usepackage{mathtools}
\usepackage{multirow}
\usepackage{algpseudocode}
\usepackage[linesnumbered,ruled,vlined]{algorithm2e}
\usepackage{setspace}
\usepackage{footmisc}
\usepackage[justification=centering]{caption}
\usepackage{dsfont}
\usepackage{bbm}
\usepackage{cases}
\usepackage{stfloats}

\theoremstyle{definition}
\newtheorem{remark}{Remark}
\newtheorem{corollary}{Corollary}

\newtheorem{theorem}{Theorem}

\newtheorem{lemma}{Lemma}

\makeatletter
\newcommand{\biggg}{\bBigg@{3}}
\newcommand{\Biggg}{\bBigg@{3.5}}
\makeatother
\usepackage{bm}
\begin{document}
\title{Downlink and Uplink NOMA-ISAC \\with Signal Alignment}

\author{Boqun Zhao, \IEEEmembership{Graduate Student Member, IEEE}, 
        Chongjun Ouyang, \IEEEmembership{Member, IEEE},\\
        Xingqi Zhang, \IEEEmembership{Member, IEEE}, and Yuanwei Liu, \IEEEmembership{Fellow, IEEE}
\vspace{-5pt}
\thanks{B. Zhao and X. Zhang are with Department of Electrical and Computer Engineering, University of Alberta, Edmonton AB, T6G 2R3, Canada (email: \{boqun1, xingqi.zhang\}@ualberta.ca).}
\thanks{C. Ouyang is with the School of Electrical and Electronic Engineering, University College Dublin, Dublin, D04 V1W8, Ireland, and also with the School of Electronic Engineering and Computer Science, Queen Mary University of London, London, E1 4NS, U.K. (e-mail: chongjun.ouyang@ucd.ie).}
\thanks{Y. Liu is with the School of Electronic Engineering and Computer Science, Queen Mary University of London, London, E1 4NS, U.K. (email: yuanwei.liu@qmul.ac.uk).}
}
\maketitle
\begin{abstract}
Integrated Sensing and Communications (ISAC) surpasses the conventional frequency-division sensing and communications (FDSAC) in terms of spectrum, energy, and hardware efficiency, with potential for greater enhancement through integration of non-orthogonal multiple access (NOMA). Leveraging these advantages, a multiple-input multiple-output NOMA-ISAC framework is proposed in this paper, in which the technique of signal alignment is adopted. The performance of the proposed framework for both downlink and uplink is analyzed. 1) The downlink ISAC is investigated under three different precoding designs: a sensing-centric (S-C) design, a communications-centric (C-C) design, and a Pareto optimal design. 2) For the uplink case, two scenarios are investigated: a S-C design and a C-C design, which vary based on the order of interference cancellation between the communication and sensing signals. In each of these scenarios, key performance metrics including sensing rate (SR), communication rate (CR), and outage probability are investigated. For a deeper understanding, the asymptotic performance of the system in the high signal-to-noise ratio (SNR) region is also explored, with a focus on the high-SNR slope and diversity order. Finally, the SR-CR rate regions achieved by ISAC and FDSAC are studied. Numerical results reveal that in both downlink and uplink cases, ISAC outperforms FDSAC in terms of sensing and communications performance and is capable of achieving a broader rate region, clearly showcasing its superiority.
\end{abstract}
\section{Introduction}
The concept of Integrated Sensing and Communications (ISAC) has sparked considerable attention from both the research community and industry due to its immense potential in facilitating the advancement of sixth-generation (6G) and forthcoming wireless networks \cite{overview2}. A salient attribute of ISAC is its capacity to concurrently utilize the same time, frequency, power, and hardware resources for both communication and sensing purposes. This stands in stark contrast to the conventional approach of Frequency-Division Sensing and Communications (FDSAC), which necessitates distinct frequency bands and infrastructure for the two functions. The efficiency of ISAC is thus expected to surpass FDSAC in terms of spectrum utilization, energy consumption, and hardware requirements \cite{overview1,LiuAn}.

In the assessment of ISAC's effectiveness, two crucial performance metrics are commonly employed from an information-theoretic perspective: sensing rate (SR) and communication rate (CR) \cite{LiuAn,Tang2019_TSP}. The SR is a measure of the system's capability to accurately estimate environmental information through sensing processes. On the other hand, the CR quantifies the system's capacity for efficient data transmission during communication operations. By analyzing these two metrics, researchers can gain valuable insights into the overall performance and efficacy of ISAC in seamlessly integrating sensing and communications (S\&C) functionalities. On this basis, to further define ISAC's fundamental performance limits and evaluate the S\&C performance tradeoffs, \cite{ouyang2022integrated} proposed a novel mutual information (MI)-based framework for ISAC, where the SR and CR are characterized by the sensing MI and the communication MI, respectively.

In recent times, there has been a significant growth in the literature concerning the performance of ISAC. The author in \cite{single_UT_1} approached the characterization of the SR-CR region in ISAC systems with a single communication user terminal (UT) from an information-theoretical perspective. Furthermore, in \cite{single_UT_2}, the trade-offs between S\&C performance in single-UT ISAC systems were discussed, adopting an estimation-theoretical viewpoint by utilizing the Cramér-Rao bound metric for sensing and the minimum mean-square error metric for communications. Utilizing the MI-based framework proposed in \cite{ouyang2022integrated}, a previous work \cite{ouyang_beamforming} investigated the fundamental performance limits of ISAC systems involving a single-antenna UT under various beamforming vector designs. Building on this, researchers have expanded their inquiries to multi-UT scenarios where each UT is equipped with a single antenna, as explored in \cite{Ouyang2022_WCL}. In this work, dirty paper coding is employed for interference management among communication users, and it is assumed that the interference from the sensing signal to the communication signal is perfectly mitigated to prioritize communications. These approaches, while effective, demand computationally intensive implementations. To overcome these challenges, a recent work \cite{ouyang_mimoisac} has investigated downlink multiple-input multiple-output (MIMO)-ISAC systems featuring multiple UTs, each equipped with multiple antennas. In this setup, space division multiple access (SDMA) is utilized to manage user interference. However, due to inherent limitations, the number of serviceable UTs is capped by the number of transmit antennas at the base station (BS).

The existing research in multi-antenna ISAC systems has made significant contributions. However, it is important to acknowledge that when the system becomes overloaded, the UTs can experience severe inter-user interference (IUI) due to the constraints imposed by limited spatial degrees of freedoms (DoFs). This interference can adversely impact the performance of the system, leading to reduced efficiency and potentially compromising the overall effectiveness. To address this issue, non-orthogonal multiple access (NOMA) can be employed to mitigate IUI and further improve the system performance. NOMA's power allocation based on individual user channel conditions effectively reduces IUI by empowering weaker channels to overcome interference from stronger ones \cite{tang_1,userparing}. Additionally, employing successive interference cancellation (SIC) at the receiver side further mitigates IUI's impact on subsequent users \cite{boqun,tang_2}. The superior spectral efficiency of NOMA allows it to accommodate more users than the conventional multiple access techniques, e.g., SDMA, making it a valuable enhancement for ISAC systems.

The concept of NOMA-ISAC has been introduced in several works \cite{xidong,NOMA_ISAC_1, NOMA_ISAC_2, NOMA_ISAC_4, NOMA_ISAC_5}. However, it is important to highlight that most of these studies have primarily focused on waveform or beamforming design aspects. In contrast, there has been limited quantitative analysis of the fundamental performance of NOMA-ISAC systems, with only a few recent works addressing this issue \cite{Chao, uplink_SIC,ouyang_globecom}. Specifically, the authors of \cite{Chao} analyzed the performance of an uplink NOMA-based Semi-ISAC network with a single-input-single-output model, considering only one pair of UTs and thus neglecting inter-pair interference (IPI). Similarly, the work in \cite{uplink_SIC} proposed an uplink NOMA-ISAC framework with single-antenna UTs, exploring the impact of different decoding orders for the S\&C signals. On the other hand, the performance of downlink NOMA-ISAC systems has been only studied in our previous work \cite{ouyang_globecom}, where a basic framework of single-input single-output NOMA-ISAC with a single UT pair is developed. Therefore, research on the performance characterization of NOMA-ISAC is still in its early stages, and the more general NOMA-ISAC within a MIMO framework involving multiple pairs of UTs remains unexplored, which provides the motivation for this paper. 

As a comprehensive extension of our previous works \cite{ouyang_beamforming,ouyang_mimoisac,ouyang_globecom}, this article studies a more general NOMA-ISAC model involving multiple UTs, each equipped with multiple antennas. Based on the aforementioned MI-based framework, we conduct a thorough performance analysis by encompassing both downlink and uplink scenarios. Moreover, to enhance system performance, we introduce signal alignment into the proposed NOMA-ISAC system, effectively leveraging excess DoFs to suppress co-channel IPI \cite{signalalignment}. Consequently, the proposed system in this paper is capable to serve more UTs than the conventional MIMO-ISAC systems \cite{ouyang_mimoisac}. The main contributions of this paper are summarized as follows:
\begin{itemize}
  \item We propose a novel MIMO-based NOMA-ISAC framework designed to cater to both downlink and uplink scenarios. The system involves a dual-functional S\&C (DFSAC) BS that efficiently serves multiple communication UTs, each equipped with multiple antennas, while concurrently sensing the targets. To overcome the challenges posed by IPI and enhance system throughput, we employ the concept of signal alignment. By doing so, we determine the appropriate detection vector for downlink transmission and the precoding vector for uplink transmission, effectively aligning signals to minimize interference and optimize performance of communications.
  \item We conduct a comprehensive analysis of the downlink ISAC performance, considering three typical scenarios: sensing-centric (S-C) design, communications-centric (C-C) design, and Pareto optimal design. Within each scenario, we derive key performance metrics, SR, CR and outage probability (OP). Notably, we also investigate the high-SNR slopes and diversity orders to gain more insights of the system performance. To provide a meaningful baseline for comparison, we evaluate the performance of downlink FDSAC and characterize the achievable SR-CR regions for both ISAC and FDSAC. In particular, we mathematically prove that the rate region achieved by ISAC entirely encompasses that achieved by FDSAC. 
  \item We conduct a comprehensive analysis of the uplink ISAC performance considering two distinct scenarios: S-C design and C-C design, each employing different interference cancellation orders for S\&C signals at the receiver. We derive the same key performance metrics as in the downlink case. We also obtain an achievable rate region of uplink ISAC by means of the time-sharing strategy. Furthermore, we compare the ISAC performance with the conventional uplink FDSAC.
  \item Numerical results are presented and demonstrate that in both uplink and downlink scenarios, the S-C design exhibits superior sensing performance, while the C-C design excels in communication performance. Remarkably, both S-C and C-C designs outperform the conventional FDSAC system. Particularly noteworthy is that the SR-CR rate regions of FDSAC are entirely encompassed within the regions of ISAC in both uplink and downlink cases. These findings underscore the significant advantages of the proposed ISAC framework in optimizing the trade-offs between S\&C performance, surpassing the capabilities of the conventional FDSAC system in various operational scenarios. 
\end{itemize}

The rest of this paper is organized as follows. Section~\ref{system} outlines the conception of the NOMA-ISAC framework, encompassing both downlink and uplink scenarios. Section~\ref{downlink} focuses on the downlink ISAC performance, presenting the results for the three designs (S-C, C-C, and Pareto optimal) and the downlink FDSAC, as well as the rate regions. In Section~\ref{uplink}, we delve into the uplink ISAC performance, introducing the results for the two different designs (S-C and C-C) and the uplink FDSAC, also characterizing the rate region. Numerical results are provided in Section~\ref{numerical}, followed by the conclusion in Section~\ref{conclusion}.

 \begin{figure} [t!]
\centering
\includegraphics[height=0.22\textwidth]{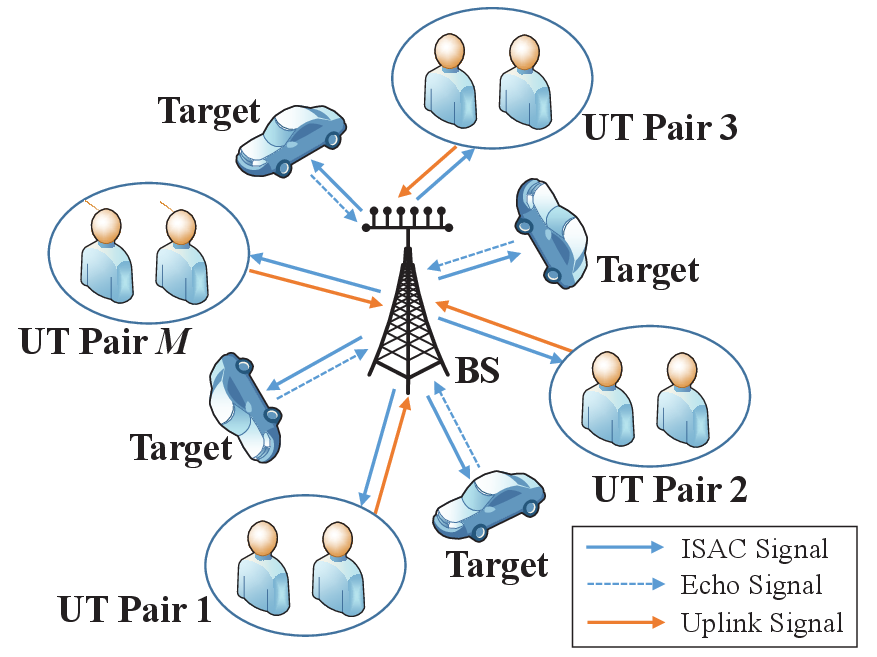}
 \caption{\!Illustration of a\! downlink/uplink NOMA-ISAC system.}
 \vspace{-10pt}
 \label{system_model}
\end{figure}

\section{System Model}\label{system}
Consider a downlink/uplink NOMA-ISAC system as depicted in {\figurename} \ref{system_model}, where a DFSAC BS is communicating with multiple UTs equipped with $N$ antennas each, while simultaneously sensing the targets in its surrounding environment. The BS operates in full duplex mode with two distinct sets of antennas, $M$ ($M\leqslant N$)\footnote{It is worth noting that the assumption of $M\leqslant N$ generally holds for some low-cost and low-power small cells where the BS has the same number of antennas as UTs, or even less \cite{antenna_num}. This assumption has been also commonly used in current literature \cite{ouyang_mimoisac,MIMO_NOMA2}.} transmit antennas and $M$ receive antennas\footnote{Regarding sensing, to mitigate information loss related to sensed targets, it is commonly stipulated that the number of receive antennas should not be less than the number of transmit antennas. Thus, in pursuit of resource efficiency, we choose to employ the minimum number of receive antennas, equalling the number of transmit antennas, while certain findings can be extended to cases involving more receive antennas.}, which are well-separated in space \cite{Tang2019_TSP}. In this paper, due to the benefits of NOMA with signal alignment, the proposed ISAC system has the capability to serve up to $2M$ UTs, doubling the capacity of the existing works \cite{ouyang_mimoisac,uplink_SIC}. This expanded capacity will be justified in the subsequent analysis. To alleviate the system load, previous studies on NOMA have proposed to pair two users for the implementation of NOMA \cite{signalalignment, boqun}. Building upon this concept, we presume the utilization of random user-pairing\footnote{We note that however sophisticated user pairing is capable of enhancing the performance of the networks considered, which is set aside for our future work.} in this study \cite{userparing}, resulting in the formation of a total of $M$ pairs. Specifically, each pair consists of a UT near the BS denoted as $m$, and the more distant UT denoted as $m^\prime$, where $m=1,\ldots,M$. For simplicity, we assume that the BS and UTs have full channel state information (CSI) and that the BS can perfectly eliminate the echo signal reflected by the UTs \cite{Ouyang2022_WCL}.
\subsection{Downlink ISAC}
Consider a DFSAC signal matrix  $\mathbf{X}=\left[{\mathbf x}_1 \ldots {\mathbf x}_L\right]\in{\mathbbmss{C}}^{M\times L}$ sent from the BS, where $L$ represents the length of the communication frame/sensing pulse. In the context of sensing, ${\mathbf x}_l\in{\mathbbmss{C}}^{M\times1}$ for $l=1,\ldots,L$ represents the sensing snapshot transmitted during the $l$th time slot. For communication purposes, ${\mathbf x}_l$ corresponds to the $l$th data symbol vector. Under the framework of MIMO-based ISAC, we can express the downlink signal matrix $\mathbf{X}$ as follows:
\begin{align}\label{dual_function_signal_matrix}
\mathbf{X}=\mathbf{P}\mathbf{S}=\mathbf{W}_{\rm{d}}\mathbf{\Xi }^{\frac{1}{2}}\mathbf{S},
\end{align}
where $\mathbf{P}=\mathbf{W}_{\rm{d}}\mathbf{\Xi }^{\frac{1}{2}}\in \mathbbmss{C} ^{M\times M}$ denotes the downlink precoding matrix at the BS, $\mathbf{W}_{\rm{d}}=\left[ \mathbf{w}_{{\rm{d}},1}\ldots\mathbf{w}_{{\rm{d}},M} \right] \in \mathbbmss{C} ^{M\times M}$ contains the normalized precoders with $\lVert{\mathbf{w}}_{{\rm{d}},m}\rVert^2=1$ for $m=1,\ldots,M$, $\mathbf{\Xi }=\mathsf{diag}\left\{ p_1,...,p_M \right\} \succcurlyeq \mathbf{0}$ is the power allocation matrix subject to the power budget $\sum_{m=1}^M{p_m\leqslant p}$, and $\mathbf{S}=\left[{\mathbf{s}}_1\ldots{\mathbf{s}}_M\right]^{\mathsf{H}}\in{\mathbbmss{C}}^{M\times L}$ contains $M$ unit-power data streams intended for the $M$ pairs. Specifically, it follows that
\begin{align}
\mathbf{s}_{m}=\sqrt{\alpha _m}\mathbf{s}_{\mathrm{d},m}+\sqrt{\alpha _{m^{\prime}}}\mathbf{s}_{\mathrm{d},m^{\prime}},
\end{align}
where $\mathbf{s}_{\mathrm{d},m}\in \mathbbmss{C} ^{L \times 1}$ and $\mathbf{s}_{\mathrm{d},m'} \in \mathbbmss{C} ^{L \times 1}$ denote the unit-power information bearing signals to be transmitted to UT $m$ and UT $m'$, respectively, and $\alpha _ m$ and $\alpha _{m'}$ denote the NOMA power allocation coefficients with $\alpha _ m+\alpha _{m'}=1$. The data streams are assumed to be independent with each other such that $L^{-1}\mathbf{S}\mathbf{S}^{\mathsf{H}}=\mathbf{I}_M$.
\subsubsection{Sensing Model}
When transmitting the signal matrix $\mathbf{X}$ for target sensing, the BS receives the reflected echo signal matrix, which can be expressed as follows: \cite{Tang2019_TSP,Ouyang2022_WCL,ouyang_mimoisac}
\begin{align}\label{reflected_echo_signal_matrix}
{\mathbf{Y}}_{\rm{s}}={\mathbf{G}}{\mathbf{X}}+{\mathbf{N}}_{\rm{s}},
\end{align}
where ${\mathbf{N}}_{\rm{s}}\in{\mathbbmss{C}}^{M\times L}$ is the additive white Gaussian noise (AWGN) matrix with each entry having zero mean and unit variance, and $\mathbf{G}=\left[{\mathbf g}_1 \ldots {\mathbf{g}}_M\right]^{\mathsf{H}}\in{\mathbbmss{C}}^{M\times M}$ represents the target response matrix with ${\mathbf{g}}_m\in{\mathbbmss{C}}^{M\times 1}$ for $m=1,\ldots,M$ characterizing the target response from the transmit antenna array to the $m$th receive antenna. The target response matrix is modeled as \cite{Tang2019_TSP,Ouyang2022_WCL}
\begin{align}
\mathbf{G}=\sum\nolimits_{k}\beta_k{\mathbf{a}}\left(\theta_k\right){\mathbf{b}}^{\mathsf{T}}\left(\theta_k\right),
\end{align}
where $\beta_k$ denotes the radar cross section (RCS) of the $k$th target, ${\mathbf{a}}\left(\theta_k\right)\in{\mathbbmss{C}}^{M\times1}$ and ${\mathbf{b}}\left(\theta_k\right)\in{\mathbbmss{C}}^{M\times1}$ are the associated receive and transmit array steering vectors, respectively, and $\theta_k$ is its direction of arrival. Assume that the BS is equipped with uniform linear arrays with half-wavelength spacing, which yields $\mathbf{a}\left( \theta _k \right) =\mathbf{b}\left( \theta _k \right) =\left[ \mathrm{e}^{\mathrm{j}\pi \left( m-1 \right) \sin \theta _k} \right] _{m=1}^{M}$. Assuming that the receive antennas at the BS are widely separated, we have $\mathbf{g}_m\sim{\mathcal{CN}}\left({\mathbf{0}},\mathbf{R}\right)$ for $m=1,\ldots,M$ and ${\mathbb{E}}\left\{{\mathbf{g}}_m{\mathbf{g}}_{m'}^{\mathsf{H}}\right\}={\mathbf{0}}$ for $m\neq m'$ \cite{Tang2019_TSP}. Following the Swerling-\uppercase\expandafter{\romannumeral1} model, we assume that the RCS is relatively constant from pulse-to-pulse with an \emph{a prior} Rayleigh distributed amplitude, in which case $\beta_k$ follows the complex Gaussian distribution ${\mathcal{CN}}(0,\sigma_k^2)$. Note that $\sigma_k^2$ is the arithmetic mean of all values of RCS of the reflecting object, and represents the average reflection strength. It is noteworthy that the statistics of the target response are primarily influenced by the statistical characteristics of these RCSs.

In contrast to the instantaneous target response matrix $\mathbf{G}$, the correlation matrix ${\mathbf{R}}\in{\mathbbmss{C}}^{M\times M}$ remains relatively stable over an extended period. Obtaining $\mathbf{R}$ is a straightforward task for the BS through long-term feedback. Therefore, for the purposes of our discussion, we assume that the BS has access to the correlation matrix $\mathbf{R}$, which is the assumption commonly made in the literature \cite{Tang2019_TSP,Ouyang2022_WCL}.

The primary objective of sensing is to extract environmental information contained in the target response $\mathbf{G}$, such as the direction and reflection coefficient of each target, from the reflected echo signal ${\mathbf{Y}}_{\rm{s}}$ \cite{Tang2019_TSP,Ouyang2022_WCL}. This extraction of information is quantified by the MI between ${\mathbf{Y}}_{\rm{s}}$ and $\mathbf{G}$ conditioned on the DFSAC signal ${\mathbf{X}}$, which is commonly referred to as the sensing MI \cite{Tang2019_TSP,ouyang2022integrated}. From an information-theoretical perspective, the sensing performance is evaluated using the SR, defined as the sensing MI per unit time \cite{ouyang2022integrated}. Besides, under our considered model, maximizing the SR is equivalent to minimizing the mean-square error (MSE) in estimating the target response $\mathbf{G}$ \cite{ouyang2022integrated,MSE}, as detailed in Appendix \ref{Appendix:0}. Assuming that each DFSAC symbol lasts for 1 unit of time, the SR is expressed as follows:
\begin{align}\label{eq_SR}
{\mathcal{R}}_{\rm{s}}=L^{-1}I\left({\mathbf{Y}}_{\rm{s}};\mathbf{G}|\mathbf{X}\right),    
\end{align}
where $I\left(X;Y|Z\right)$ represents the MI between random variables $X$ and $Y$ conditioned on $Z$.

\subsubsection{Communication Model}
By transmitting $\mathbf{X}$ to the UTs, the received signal matrix at UT $m$ is given by
\begin{align}
\mathbf{Y}_{\mathrm{c},m}=&\sqrt{\eta_m^{-1}p_m}\mathbf{H}_{\mathrm{d},m}\mathbf{w}_{{\rm{d}},m}\mathbf{s}_{m}^{\mathsf{H}} \notag\\
&+{\sum}_{i\ne m}{\sqrt{\eta_m^{-1}p_i}\mathbf{H}_{\mathrm{d},m}\mathbf{w}_{{\rm{d}},i}}\mathbf{s}_{i}^{\mathsf{H}} +\mathbf{N}_{m},
\end{align}
where $\eta _m$ denotes the large-scale path loss for UT $m$, ${\mathbf{H}}_{\mathrm{d},m}\in{\mathbbmss{C}}^{N\times M}$ is the downlink communication channel matrix with each element representing independent and identically distributed (i.i.d.) Rayleigh fading channel gains, and ${\mathbf{N}}_{m}\in{\mathbbmss{C}}^{N\times L}$ is the AWGN matrix with each entry having zero mean and unit variance.

After receiving $\mathbf{Y}_{{\rm{c}},m}$, the UT adopts a detection vector (equalizer) $\mathbf{v}_{\mathrm{d},m}\in{\mathbbmss{C}}^{N\times 1}$, and thus the UT's observation can be rewritten as follows:
\begin{align}
\mathbf{v}_{\mathrm{d},m}^{\mathsf{H}}\!\mathbf{Y}_{\mathrm{c},m}\!&=\sqrt{\eta_m^{-1}p_m}\mathbf{v}_{\mathrm{d},m}^{\mathsf{H}}\mathbf{H}_{\mathrm{d},m}\mathbf{w}_{{\rm{d}},m}\mathbf{s}_{m}^{\mathsf{H}} \notag\\
&\!\!\!\!+\underset{\mathrm{IPI}}{\underbrace{{\sum}_{i\ne m}\!{\sqrt{\eta_m^{-1}p_i}\mathbf{v}_{\mathrm{d},m}^{\mathsf{H}}\mathbf{H}_{\mathrm{d},m}\mathbf{w}_{{\rm{d}},i}}\mathbf{s}_{i}^{\mathsf{H}} }}\!+\!\mathbf{v}_{\mathrm{d},m}^{\mathsf{H}}\mathbf{N}_{m}.
\end{align}
To eliminate the IPI, it's necessary to fulfill the conditions $\mathbf{v}_{\mathrm{d},i}^{\mathsf{H}}\mathbf{H}_{\mathrm{d},i}\mathbf{w}_{\mathrm{d},m}=0$ and $\mathbf{v}_{\mathrm{d},i'}^{\mathsf{H}}\mathbf{H}_{\mathrm{d},i'}\mathbf{w}_{\mathrm{d},m}=0$ for $i=1,..,m-1,m+1,...,M$. In this context, from the perspective of designing $\mathbf{w}_{\mathrm{d},m}$, a total of $2(M-1)$ equations must be satisfied, while only $M$ variables are available. Consequently, unless the total number of UT pairs is no greater than $\frac{M}{2}+1$, it's not feasible for a non-zero vector $\mathbf{w}_{\mathrm{d},m}$ to exist \cite{signalalignment}. However, reducing the number of pairs will inevitably result in a reduction of the overall system throughput.

In order to overcome this problem, the concept of signal alignment can be applied, where the detection vectors are designed to satisfy the constraint  $\mathbf{v}_{\mathrm{d},m}^{\mathsf{H}}\mathbf{H}_{\mathrm{d},m}=\mathbf{v}_{\mathrm{d},m'}^{\mathsf{H}}\mathbf{H}_{\mathrm{d},m'}$ \cite{signalalignment}. By applying the signal alignment, the channels of the two UTs within the same pair are projected to a common direction, effectively reducing the dimension of the received signals, i.e., only $M-1$ equations need to be satisfied. This alignment enables the proposed system to accommodate $2M$ UTs, providing the potential to serve a larger number of users. The designs of such $\mathbf{v}_{\mathrm{d},m}$ and $\mathbf{v}_{\mathrm{d},m'}$ as well as the preocoder $\mathbf{W}_{\rm{d}}$ will be detailed in the next section.

Based on the above analysis, the effective channel gains at UT $m$ and $m'$ are expressed as $g_m=\eta _m^{-1}\left| \mathbf{v}_{\mathrm{d},m}^{\mathsf{H}}\mathbf{H}_{\mathrm{d},m}\mathbf{w}_{\mathrm{d},m} \right|^2$ and $g_{m^{\prime}}=\eta _{m^{\prime}}^{-1}\left| \mathbf{v}_{\mathrm{d},m'}^{\mathsf{H}}\mathbf{H}_{\mathrm{d},m'}\mathbf{w}_{\mathrm{d},m} \right|^2$, respectively. Because the path loss is positively correlated with the distance to the BS, i.e., $\eta _m^{-1}>\eta _{m^{\prime}}^{-1}$, and $\mathbf{v}_{\mathrm{d},m}^{\mathsf{H}}\mathbf{H}_{\mathrm{d},m}=\mathbf{v}_{\mathrm{d},m'}^{\mathsf{H}}\mathbf{H}_{\mathrm{d},m'}$, we have $g_m>g_{m'}$. Therefore, according to the principle of NOMA, the two UTs from the same pair are ordered without any ambiguity, i.e., $\alpha _m<\alpha _{m'}$. In this context, UT $m$ performs the SIC process by first eliminating the message of $m'$ before decoding its own signals, while $m'$ directly decode its own signals. We assume having fixed power allocation sharing $\{\alpha_m,\alpha_{m'}\}$ in each UT pair, but optimal power sharing strategies are capable of further enhancing the performance of the networks considered, which is beyond the scope of this paper. As a result, the SNR of UT $m$ and the signal-to-interference-plus-noise ratio (SINR) of UT $m'$ are, respectively, given by
\begin{align}
\gamma _{\mathrm{d},m}&=\frac{\alpha _m\eta _{m}^{-1}p_m\left| \mathbf{v}_{\mathrm{d},m}^{\mathsf{H}}\mathbf{H}_{\mathrm{d},m}\mathbf{w}_{\mathrm{d},m} \right|^2}{\left\| \mathbf{v}_{\mathrm{d},m} \right\| ^2}, \\
\gamma _{\mathrm{d},m^{\prime}}&=\frac{\alpha _{m^{\prime}}\eta _{m^{\prime}}^{-1}p_m\left| \mathbf{v}_{\mathrm{d},m'}^{\mathsf{H}}\mathbf{H}_{\mathrm{d},m'}\mathbf{w}_{\mathrm{d},m} \right|^2}{\alpha _m\eta _{m'}^{-1}p_m\left| \mathbf{v}_{\mathrm{d},m'}^{\mathsf{H}}\mathbf{H}_{\mathrm{d},m'}\mathbf{w}_{\mathrm{d},m} \right|^2+\left\| \mathbf{v}_{\mathrm{d},m^{\prime}} \right\| ^2}.
\end{align}

\begin{figure*}[hb]
\hrulefill
\begin{align}
\mathbf{Y}_{\mathrm{BS}}&=\underset{\mathbf{Y}_{\mathrm{BS}}^{\rm{c}}}{\underbrace{\sum\nolimits_{m=1}^M{\left( \sqrt{\alpha _m\eta _{m}^{-1}p_{\rm{c}}}\mathbf{H}_{\mathrm{u},m}\mathbf{w}_{{\rm{u}},m}\mathbf{s}_{\mathrm{u},m}^{\mathsf{H}}+\sqrt{\alpha _{m'}\eta _{m'}^{-1} p_{\rm{c}}}\mathbf{H}_{\mathrm{u},m^{\prime}}\mathbf{w}_{{\rm{u}},m^{\prime}}\mathbf{s}_{\mathrm{u},m^{\prime}}^{\mathsf{H}} \right)}}}+\mathbf{GX}_{\rm{s}}+\mathbf{N}_{\mathrm{u}}. \label{Uplink_DFSAC_Signal} \\
\mathbf{v}_{\mathrm{u},m}^{\mathsf{H}}\mathbf{Y}_{\mathrm{BS}}^{\rm{c}}&=\left( \sqrt{\alpha _m\eta _{m}^{-1}p_{\rm{c}}}\mathbf{v}_{\mathrm{u},m}^\mathsf{H}\mathbf{H}_{\mathrm{u},m}\mathbf{w}_{{\rm{u}},m}\mathbf{s}_{\mathrm{u},m}^{\mathsf{H}}+\sqrt{\alpha _{m'}\eta _{m'}^{-1}p_{\rm{c}}}\mathbf{v}_{\mathrm{u},m}^\mathsf{H}\mathbf{H}_{\mathrm{u},m'}\mathbf{w}_{{\rm{u}},m'}\mathbf{s}_{\rm{u},m^{\prime}}^{H} \right) \nonumber\\
&~~~~+\underset{\mathrm{IPI}}{\underbrace{\sum\nolimits_{i\ne m}{\left( \sqrt{\alpha _i\eta _{i}^{-1}p_{\rm{c}}}\mathbf{v}_{\mathrm{u},m}^\mathsf{H}\mathbf{H}_{\mathrm{u},i}\mathbf{w}_{{\rm{u}},i}\mathbf{s}_{\mathrm{u},i}^{H}+\sqrt{\alpha _{i'}\eta _{i'}^{-1}p_{\rm{c}}}\mathbf{v}_{\mathrm{u},m}^\mathsf{H}\mathbf{H}_{\mathrm{u},i'}\mathbf{w}_{{\rm{u}},i'}\mathbf{s}_{\mathrm{u},i^{\prime}}^{\mathsf{H}}\right)}}}.\label{detected_signal}
\end{align}
\setcounter{equation}{11}
\end{figure*}
Given the downlink NOMA-ISAC framework, our objective is to analyze its S\&C performance by evaluating the CR and SR. Both SR and CR are influenced by the precoding matrix $\mathbf{P}$, giving rise to a multi-object optimization problem. Solving this problem is generally challenging due to the difficulty in finding a precoding policy that can maximize both CR and SR simultaneously. To address this challenge, we identify three limiting cases to unveil the performance bounds for the ISAC system under our settings in Section \ref{downlink}. The first case aims to maximize CR (C-C design), the second case aims to maximize SR (S-C design), and the third case seeks all achievable SR-CR pairs, essentially finding the Pareto outer boundary of the SR-CR region (Pareto optimal design). These considerations are fundamental operations for handling multi-object optimization problems. The C-C design prioritizes maximizing CR without considering SR, leading to poor sensing performance but establishing the upper bound of communication performance. The S-C design prioritizes maximizing SR without considering CR, leading to the upper bound of sensing performance but poor communication performance. The Pareto optimal design considers trade-offs between communications and sensing, though its characterization is more computationally intensive than the other two designs.
\subsection{Uplink ISAC}
For the uplink case, the signal observed by the BS is given as \eqref{Uplink_DFSAC_Signal}, shown at the bottom of the next page,
where $\mathbf{Y}_{\mathrm{BS}}^{\rm{c}}\in{\mathbbmss{C}}^{M \times L}$ is the communication signal, $p_{\rm{c}}$ reflects the power constraint of each UT pair, $\mathbf{H}_{\mathrm{u},m}\in{\mathbbmss{C}}^{M \times N}$ and $\mathbf{H}_{\mathrm{u},m'}\in{\mathbbmss{C}}^{M \times N}$ denote the uplink communication channel matrices of UT $m$ and UT $m'$, respectively, with elements representing i.i.d. Rayleigh fading channel gains, $\mathbf{w}_{{\rm{u}},m}\in{\mathbbmss{C}}^{N \times 1}$ and $\mathbf{w}_{{\rm{u}},m^{\prime}}\in{\mathbbmss{C}}^{N \times 1}$ denote the precoding vectors for UT $m$ and UT $m'$, respectively, $\mathbf{s}_{\mathrm{u},m}=\left[ s_{\mathrm{u},m,1}\ldots s_{\mathrm{u},m,L} \right]^\mathsf{H} \in{\mathbbmss{C}}^{ L\times 1}$ and $\mathbf{s}_{\mathrm{u},m'}=\left[ s_{\mathrm{u},m',1}\ldots s_{\mathrm{u},m',L} \right]^\mathsf{H} \in{\mathbbmss{C}}^{ L\times 1}$ are the unit-power communication messages sent by $m$ and $m'$ to the BS, respectively, with the symbols sent at different time slots being uncorrelated, i.e., $\mathbb{E} \left\{ \mathbf{s}_{\mathrm{u},m}\mathbf{s}_{\mathrm{u},m}^{\mathsf{H}} \right\} =\mathbb{E} \left\{ \mathbf{s}_{\mathrm{u},m^{\prime}}\mathbf{s}_{\mathrm{u},m^{\prime}}^{\mathsf{H}} \right\} =\mathbf{I}_L$, $\mathbf{X}_{\rm{s}}=\left[ \mathbf{x}_{{\rm{s}},1}, \ldots ,\mathbf{x}_{{\rm{s}},L} \right]\in{\mathbbmss{C}}^{ M\times L} $ denotes the uplink sensing signal with $\mathbf{x}_{{\rm{s}},l}\in{\mathbbmss{C}}^{ M\times 1}$ being the waveform at the $l$th time slot, and $\mathbf{N}_{\mathrm{u}}=\left[\mathbf{n}_{{\rm{u}},1}, \ldots, \mathbf{n}_{{\rm{u}},L}\right]\in{\mathbbmss{C}}^{M \times L}$ is the AWGN matrix with each entry having zero mean and unit variance. 

A Two-Stage SIC Process: After receiving the above signal, the BS is tasked with decoding both the communication signal's data information and the environmental information contained in the target response $\mathbf{G}$. To address both inter-functionality interference (IFI) and IUI, a two-stage SIC process can be employed \cite[Fig. 2]{uplink_SIC}. The outer-stage SIC focuses on handling the IFI, while the inner-stage SIC tackles the IUI. In the context of the outer-stage SIC, we consider two SIC orders. In the first SIC order, the BS initially detects the communication signal by treating the sensing signal as interference \cite{Ouyang2022_WCL,GaussianNoise}. Subsequently, the detected communication signal is subtracted from the superposed signal, with the remaining portion utilized for sensing the target response. In the second SIC order, the BS first senses the target response signal by treating the communication signal as interference. Then, the sensing signal is subtracted, leaving the remaining part for detecting the communication signal. It is evident that the first SIC order offers superior sensing performance, while the second SIC order yields improved communication performance. Therefore, we refer to these two SIC orders as the S-C design and C-C design, respectively, for the uplink ISAC. Within the inner-stage SIC, the messages for each UT are decoded. As the sum rate of each user group remains constant in uplink NOMA, regardless of the decoding order \cite{signalalignment}, we assume that the power allocation coefficients and the decoding order are identical to the downlink case, i.e., UT $m'$ is decoded first in the inner-stage SIC process. 

Importantly, prior to performing the inner-stage SIC, it's imperative to initially mitigate the IPI within the communication signal. Therefore, in the following discussion, with the exclusion of the sensing signal and noise, we focus on the communication signal to introduce the process of IPI elimination through signal alignment.
By applying the detection vector $\mathbf{v}_{\mathrm{u},m} \in {\mathbbmss{C}}^{M \times 1}$ to detect the message from the $m$th UT pair, the communication signal is expressed as \eqref{detected_signal}, shown at the bottom of this page.

To remove the IPI, we need to satisfy $\mathbf{v}_{\mathrm{u},m}^\mathsf{H}\mathbf{H}_{\mathrm{u},i}\mathbf{w}_{{\rm{u}},i}=\mathbf{v}_{\mathrm{u},m}^\mathsf{H}\mathbf{H}_{\mathrm{u},i'}\mathbf{w}_{{\rm{u}},i'}=0$ for $\forall i\ne m$. By leveraging the concept of signal alignment, we impose the constraint $\mathbf{H}_{\mathrm{u},m}\mathbf{w}_{{\rm{u}},m}=\mathbf{H}_{\mathrm{u},m'}\mathbf{w}_{{\rm{u}},m'}$ on the precoding vectors. Specifically, we can design the precoding vectors as follows:
\begin{align}\label{signalalignment}
\left[ \begin{matrix}
	\mathbf{w}_{\mathrm{u},m}^{\mathsf{H}}&		\mathbf{w}_{\mathrm{u},m'}^{\mathsf{H}}\\
\end{matrix} \right] ^{\mathsf{H}}=\mathbf{\Psi }_m\mathbf{r}_m,
\end{align}
where the matrix $\bm{\Psi  }_{m}$ has dimensions $2N\times (2N-M)$ and comprises the $2N-M$ right singular vectors of $\left[ \begin{matrix}
\mathbf{H}_{\mathrm{u},m}& -\mathbf{H}_{\mathrm{u},m'}\
\end{matrix} \right] $ corresponding to its zero singular values, and $\mathbf{r}_{m}$ is a $\left( 2N-M \right) \times 1$ vector which can be arbitrarily chosen and is used to constrain the total transmission power of each UT pair. Following the principle in \cite{signalalignment}, we set $\left\| \mathbf{r}_{m} \right\|^2=2$ to satisfy the power constraint:
\begin{align}
&p_{\mathrm{c}}\alpha _m\left\| \mathbf{w}_{\mathrm{u},m} \right\| ^2+p_{\mathrm{c}}\alpha _{m^{\prime}}\left\| \mathbf{w}_{\mathrm{u},m^{\prime}} \right\| ^2 \notag\\
&~\leqslant p_{\mathrm{c}}\max \left( \alpha _m,\alpha _{m^{\prime}} \right) \left( \left\| \mathbf{w}_{\mathrm{u},m} \right\| ^2+\left\| \mathbf{w}_{\mathrm{u},m^{\prime}} \right\| ^2 \right) \leqslant 2p_{\mathrm{c}}.   
\end{align}
By defining $\mathbf{Q}\triangleq \left[ \mathbf{H}_{\mathrm{u},1}\mathbf{w}_{{\rm{u}},1},\ldots,\mathbf{H}_{\mathrm{u},M}\mathbf{w}_{{\rm{u}},M} \right] ^{\mathsf{H}} \in \mathbbmss{C}^{M\times M}$, the normalized zero-forcing-based detection matrix is designed as follows:
\begin{align}
 \mathbf{V}_{\rm{u}}=\left[ \mathbf{v}_{\mathrm{u},1},\ldots,\mathbf{v}_{\mathrm{u},M} \right] =\mathbf{Q}^{-1}\mathbf{D},   
\end{align}
where $\mathbf{D}=\mathsf{diag}\left\{ \frac{1}{\sqrt{\left( \mathbf{Q}^{-\mathsf{H}}\mathbf{Q}^{-1} \right) _{1,1}}},\ldots,\frac{1}{\sqrt{\left( \mathbf{Q}^{-\mathsf{H}}\mathbf{Q}^{-1} \right) _{M,M}}} \right\} $ is a diagonal matrix to ensure $\left\|\mathbf{v}_{\mathrm{u},m}\right\|^2=1$ for $m=1,\ldots,M$, and $\left( \mathbf{A} \right) _{m,m}$ refers to the $m$th element on the diagonal of the matrix $\mathbf{A}$. Therefore, the detected signal from the $m$th UT pair in \eqref{detected_signal} can be rewritten as follows:
 \begin{align}
\mathbf{v}_{\mathrm{u},m}^{\mathsf{H}}\mathbf{Y}_{\mathrm{BS}}^{\rm{c}}=\frac{\sqrt{\alpha _m\eta _{m}^{-1}p_{\rm{c}}}\mathbf{s}_{\mathrm{u},m}^{\mathsf{H}}+\sqrt{\alpha _{m'}\eta _{m'}^{-1}p_{\rm{c}}}\mathbf{s}_{u,m^{\prime}}^{H}}{\sqrt{\left( \mathbf{Q}^{-\mathsf{H}}\mathbf{Q}^{-1} \right) _{m,m}}}.
 \end{align}

Having elucidated the fundamental model of the uplink ISAC, the performance of the two scenarios, the S-C design and the C-C design, will be comprehensively analyzed in Section~\ref{uplink}.
\section{Downlink Performance} \label{downlink}
This section introduces three scenarios for the downlink ISAC: S-C design, C-C design, and Pareto optimal design. In each scenario, the SR, CR, and outage performance are investigated using the concept of signal alignment. The performance of FDSAC is provided as the baseline. Furthermore, the CR-SR regions achieved by ISAC and FDSAC are characterized.
\subsection{Sensing-Centric Design}
\subsubsection{Performance of Sensing}
Under the S-C design, the precoding matrix $\mathbf{P}$ is set to maximize the downlink SR ${\mathcal{R}}_{\rm{d},\rm{s}}$. To proceed, we characterize the SR as follows.
\begin{lemma}\label{SR_Basic_Lemma}
For a given $\mathbf{P}$, ${\mathcal{R}}_{\rm{d},\rm{s}}$ can be calculated as 
\begin{align}
{\mathcal{R}}_{\rm{d},\rm{s}}=L^{-1}M\log_2\det\left({\mathbf{I}}_M+L{\mathbf{P}}^{\mathsf{H}}{\mathbf{R}}{\mathbf{P}}\right).
\end{align}
\end{lemma}
\begin{IEEEproof}
Please refer to Appendix~\ref{Appendix:A}.
\end{IEEEproof}
Under the S-C design, the precoding matrix satisfies
\begin{align}
\mathbf{P}_{\rm{s}}=\argmax\nolimits_{\mathsf{tr}( \mathbf{PP }^{\mathsf{H}} ) \le p}{\log_2\det\left({\mathbf{I}}_M+L{\mathbf{P}}^{\mathsf{H}}{\mathbf{R}}{\mathbf{P}}\right)}.
\end{align}
For analytical tractability, we assume that ${\mathbf{R}}\succ{\mathbf{0}}$. The following theorem provides the exact expression of SR achieved by the S-C design as well as its high-SNR approximation.
\begin{theorem}\label{Sensing_Rate_S_C_Theorem}
In the S-C design, the maximum downlink SR is given by
\begin{align} \label{Sensing_Rate_S_C_exact}
\mathcal{R} _{\mathrm{d},\mathrm{s}}^{\mathrm{s}}=ML^{-1}\sum\nolimits_{m=1}^M{\log _2\left( 1+L\lambda _ms_{m}^{\star} \right)},
\end{align}
where $\left\{\lambda_m>0\right\}_{m=1}^{M}$ are the eigenvalues of matrix ${\mathbf{R}}$ and $s_{m}^{\star}=\max\left\{0,\frac{1}{\nu}-\frac{1}{L\lambda_m}\right\}$ with $\sum_{m=1}^{M}\max\left\{0,\frac{1}{\nu}-\frac{1}{L\lambda_m}\right\}=p$. The maximum SR is achieved when ${\mathbf{P}}{\mathbf{P}}^{\mathsf{H}}=\mathbf{W}_{\mathrm{d}}\mathbf{\Xi W}_{\mathrm{d}}^{\mathsf{H}}={\mathbf{U}}{\bm\Delta}_{\rm{s}}{\mathbf{U}}^{\mathsf{H}}$, where ${\mathbf{U}}{\mathsf{diag}}\left\{\lambda_1,\ldots,\lambda_{M}\right\}{\mathbf{U}}^{\mathsf{H}}$ represents the eigendecomposition (ED) of ${\mathbf{R}}$ and ${\bm\Delta}_{\rm{s}}={\mathsf{diag}}\left\{s_1^{\star},\ldots,s_M^{\star}\right\}$. When $p\rightarrow\infty$, the asymptotic SR satisfies
\begin{align}\label{Sensing_Rate_S_C_Asymptotic}
\mathcal{R} _{\mathrm{d},\mathrm{s}}^{\mathrm{s},\infty}
\approx\frac{M^2}{L}\left(\log_2{p}+\frac{1}{M}\sum\nolimits_{m=1}^{M}\log_2\left(\frac{L\lambda_m}{M}\right)\right).
\end{align}
\end{theorem}
\begin{IEEEproof}
Please refer to Appendix \ref{Proof_Sensing_Rate_S_C_Theorem}.
\end{IEEEproof}
To provide more insights, we investigate the high-SNR slope based on the asymptotic SR presented in \eqref{Sensing_Rate_S_C_Asymptotic}, which is defined as follows: \cite{mimo}
\begin{align}
\mathcal{S} _{\mathrm{d},\mathrm{s}}^{\mathrm{s}}=\underset{p\rightarrow \infty}{\lim}\frac{\mathcal{R} _{\mathrm{d},\mathrm{s}}^{\mathrm{s},\infty}\left( p \right)}{\log _2p}. 
\end{align}
The high-SNR slope is a commonly used performance metric that characterizes the rate as a function of the transmit power, at high SNR, on a log scale. In the context of multi-antenna communications, it quantifies the spatial DoFs, indicating the available spatial dimensions.
\begin{remark}
The expression in \eqref{Sensing_Rate_S_C_Asymptotic} indicates that the high-SNR slope of the SR achieved under the S-C design is $\frac{M^2}{L}$.
\end{remark}
Under the S-C design, it is important to note that the normalized precoder is chosen as $\mathbf{W}_{\rm{d}}=\mathbf{U}$. Since $\mathbf{U}$ is a unitary matrix, to apply the signal alignment, we can design the normalized precoding vector as $\mathbf{w}_{\mathrm{d},\mathrm{m}}=\frac{\mathbf{H}_{\mathrm{d},m}^{\mathsf{H}}\mathbf{v}_{\mathrm{d},m}}{\left\| \mathbf{H}_{\mathrm{d},m}^{\mathsf{H}}\mathbf{v}_{\mathrm{d},m} \right\|}=\frac{\mathbf{H}_{\mathrm{d},m'}^{\mathsf{H}}\mathbf{v}_{\mathrm{d},m'}}{\left\| \mathbf{H}_{\mathrm{d},m'}^{\mathsf{H}}\mathbf{v}_{\mathrm{d},m'} \right\|}=\mathbf{u}_m$, where $\mathbf{u}_m$ represents the $m$th column of $\mathbf{U}$ with $\left\| \mathbf{u}_m \right\| =1$. In this case, the detection vectors can be found by solving the following equations:
\begin{align}\label{equations}
\mathbf{H}_{\mathrm{d},m}^{\mathsf{H}}\mathbf{v}_{\mathrm{d},m}=\mathbf{u}_m, \
\mathbf{H}_{\mathrm{d},m'}^{\mathsf{H}}\mathbf{v}_{\mathrm{d},m'}=\mathbf{u}_m.
\end{align}
Because $\mathbf{H}_{\mathrm{d},m}^{\mathsf{H}}, \mathbf{H}_{\mathrm{d},m'}^{\mathsf{H}} \in \mathbbmss{C}^{M\times N}$ are row full rank matrices, the equations are underdetermined and have infinitely many solutions. Hence, $\mathbf{v}_{\mathrm{d},m}$ and $\mathbf{v}_{\mathrm{d},m'}$ can be obtained as arbitrary solutions of \eqref{equations}, respectively. In particular, when $N=M$, $\mathbf{v}_{\mathrm{d},m}$ and $\mathbf{v}_{\mathrm{d},m'}$ can be uniquely determined, i.e., $\mathbf{v}_{\mathrm{d},m}=\mathbf{H}_{\mathrm{d},m}^{-\mathsf{H}} \mathbf{u}_m$ and $\mathbf{v}_{\mathrm{d},m'}=\mathbf{H}_{\mathrm{d},m'}^{-\mathsf{H}} \mathbf{u}_m$.
\subsubsection{Performance of Communications}\label{Section: Downlink S_C Performance of Communications}
The communication performance will be evaluated in terms of CRs and outage performance. Firstly, we focus on CRs. The CRs of UT $m$ and $m'$ under the S-C design are defined as follows: 
\begin{align}
\overline{\mathcal{R} }_{\mathrm{d},m}^{\mathrm{s}}=\log _2(1+\gamma _{\mathrm{d},m}^{\mathrm{s}}),\ \overline{\mathcal{R} }_{\mathrm{d},m'}^{\mathrm{s}}=\log _2(1+\gamma _{\mathrm{d},m'}^{\mathrm{s}}) ,
\end{align}  
where $\gamma _{\mathrm{d},m}^{\mathrm{s}}=\frac{\alpha _m\eta _{m}^{-1}s_{m}^{\star}}{\left\| \mathbf{v}_{\mathrm{d},m} \right\|^2}$ and $\gamma _{\mathrm{d},m'}^{\mathrm{s}}=\frac{\alpha _{m^{\prime}}\eta _{m^{\prime}}^{-1}s_{m}^{\star}}{\alpha _m\eta _{m'}^{-1}s_{m}^{\star}+\left\| \mathbf{v}_{\mathrm{d},m'} \right\|^2}$. We use the ergodic CRs (ECRs), i.e., $\mathcal{R} _{\mathrm{d},m}^{\mathrm{s}}=\mathbb{E} \{ \overline{\mathcal{R} }_{\mathrm{d},m}^{\mathrm{s}} \}$, $\mathcal{R} _{\mathrm{d},m'}^{\mathrm{s}}=\mathbb{E} \{ \overline{\mathcal{R} }_{\mathrm{d},m'}^{\mathrm{s}} \}$, and $\mathcal{R} _{\mathrm{d},\mathrm{c}}^{\mathrm{s}}=\sum_{m=1}^{M}(\mathcal{R} _{\mathrm{d},m}^{\mathrm{s}}+\mathcal{R} _{\mathrm{d},m'}^{\mathrm{s}})$, to assess the communication performance. 

When $N>M$, finding analytical expressions of $\mathbf{v}_{\mathrm{d},m}$ and $\mathbf{v}_{\mathrm{d},m'}$ is a challenging task, which makes the exact analysis of $\mathcal{R} _{\mathrm{d},m}^{\mathrm{s}}$ and $ \mathcal{R} _{\mathrm{d},m'}^{\mathrm{s}}$ intractable. As a compromise, we present upper bounds of $\mathcal{R} _{\mathrm{d},m}^{\mathrm{s}}$ and $ \mathcal{R} _{\mathrm{d},m'}^{\mathrm{s}}$ for the case of $N>M$. By contrast, when $N=M$, the analytical expressions of $\mathbf{v}_{\mathrm{d},m}$ and $\mathbf{v}_{\mathrm{d},m'}$ exist, and thus the exact closed-form expressions of $\mathcal{R} _{\mathrm{d},m}^{\mathrm{s}}$ and $ \mathcal{R} _{\mathrm{d},m'}^{\mathrm{s}}$ are available. The analytical results are summarized in the following Lemma.
\begin{lemma}\label{SC_ECR}
The upper bounds for $\mathcal{R} _{\mathrm{d},m}^{\mathrm{s}}$ and $ \mathcal{R} _{\mathrm{d},m'}^{\mathrm{s}}$ follow
\begin{subequations}
\begin{align}
\tilde{\mathcal{R}}_{\mathrm{d},m}^{\mathrm{s}}&=\int_0^{\infty}{\log _2\left( 1+\alpha _m\eta _{m}^{-1}s_{m}^{\star}x \right) f_{\gamma}\left( x \right) dx},\\
\tilde{\mathcal{R}}_{\mathrm{d},m'}^{\mathrm{s}}&=\int_0^{\infty}{\log _2\left( 1+\frac{\alpha _{m^{\prime}}\eta _{m^{\prime}}^{-1}s_{m}^{\star}x}{\alpha _m\eta _{m^{\prime}}^{-1}s_{m}^{\star}x+1} \right) f_{\gamma}\left( x \right) dx},
\end{align}
\end{subequations}
where $f_{\gamma}\left( x \right)$ is the probability density function (PDF) of the largest eigenvalue of the complex Wishart matrix given in \cite[Eq. (17)]{wishart}. When $N=M$, the closed-form expressions of the ECRs for UT $m$ and UT $m'$ are derived as follows:
\begin{subequations}
\begin{align}
\mathcal{R} _{\mathrm{d},m}^{\mathrm{s}}&=\frac{1}{\ln 2}\exp \left( \frac{1}{\alpha _m\eta _{m}^{-1}s_{m}^{\star}} \right) \mathrm{Ei}\left( -\frac{1}{\alpha _m\eta _{m}^{-1}s_{m}^{\star}} \right) , \label{SC_CR_M} \\
\mathcal{R} _{\mathrm{d},m'}^{\mathrm{s}}&=\frac{1}{\ln 2}\exp \left(\frac{1}{\eta _{m^{\prime}}^{-1}s_{m}^{\star}}\right)\mathrm{Ei}\left( -\frac{1}{ \eta _{m^{\prime}}^{-1}s_{m}^{\star}} \right) \notag \\
&-\frac{1}{\ln 2}\exp \left(\frac{1}{\alpha _m\eta _{m^{\prime}}^{-1}s_{m}^{\star}}\right)\mathrm{Ei}\left( -\frac{1}{\alpha _m\eta _{m^{\prime}}^{-1}s_{m}^{\star}} \right),\label{SC_CR_M'}
\end{align}
\end{subequations}
where $\mathrm{Ei}\left( x \right) =\int_{-\infty}^x{\frac{\exp \left( t \right)}{t}dt}$ denotes the exponential integral.
\end{lemma}
\begin{IEEEproof}
Please refer to Appendix~\ref{Proof_SC_ECR_Lemma}.
\end{IEEEproof}
\begin{theorem}\label{SC_sum_ECR}
Based on the results in Lemma~\ref{SC_ECR}, an upper bound of the downlink sum ECR under the S-C design is given by
\begin{align}
\tilde{\mathcal{R}}_{\mathrm{d},\mathrm{c}}^{\mathrm{s}}=\sum\nolimits_{m=1}^M{\tilde{\mathcal{R}}_{\mathrm{d},m}^{\mathrm{s}}+\tilde{\mathcal{R}}_{\mathrm{d},m^{\prime}}^{\mathrm{s}}}.
\end{align}    
Particularly, when $N=M$, the exact closed-form expression for the sum ECR is derived as
\begin{align}
\mathcal{R} _{\mathrm{d},\mathrm{c}}^{\mathrm{s}}=\sum\nolimits_{m=1}^M{\mathcal{R} _{\mathrm{d},m}^{\mathrm{s}}+\mathcal{R} _{\mathrm{d},m^{\prime}}^{\mathrm{s}}}.
\end{align}
\end{theorem}
\begin{corollary} \label{SC_ECR_infinity}
When $p\rightarrow \infty $, under the S-C design, the upper bound of the sum ECR satisfies
\begin{align} \label{SC_CR_asy_upper}
\tilde{\mathcal{R}}_{\mathrm{d},\mathrm{c}}^{\mathrm{s},\infty}\approx &M\left( \log _2p-\log _2M+\int_0^{\infty}{\log _2xf_{\gamma}\left( x \right) dx} \right) \notag\\
&+\sum\nolimits_{m=1}^M{\log _2\eta _{m}^{-1}}.
\end{align}
In the scenario where $N=M$, the asymptotic sum ECR in the high-SNR regime reads
\begin{align}\label{SC_CR_asy}
\mathcal{R} _{\mathrm{d},\mathrm{c}}^{\mathrm{s},\infty}\!\approx \!M\!\left(\! \log _2p\!-\!\log _2M\!-\!\frac{\mathcal{C}}{\ln 2} \right) \!+\!\sum\nolimits_{m=1}^M{\!\log _2\eta _{m}^{-1}},
\end{align}
where $\mathcal{C}$ denotes the Euler constant \cite{integral}.
\end{corollary}
\begin{IEEEproof}
When $p\rightarrow \infty$, we have $s_{m}^{\star}\approx \frac{p}{M}$. With the aid of ${\lim}_{x\rightarrow \infty}\log _2\left( 1+x \right) \approx \log _2x$ and \cite[Eq. (4.331.1)]{integral}, the results can be derived.
\end{IEEEproof}
\begin{remark} \label{SC_slope}
The results in \eqref{SC_CR_asy_upper} and \eqref{SC_CR_asy} suggest that the high-SNR slope of the sum ECR under the S-C design is $M$.
\end{remark}
Turn now to the outage performance. The downlink OPs of UT $m$ and $m'$ under the S-C design are defined as follows:
\begin{align}
\mathcal{P} _{\mathrm{d},m}^{\mathrm{s}}&=1-\mathrm{Pr}\left( \overline{\mathcal{R}}_{\mathrm{d},m}^{\mathrm{s}}>\mathcal{R} _{\mathrm{d},m}^{\mathrm{t}}, \overline{\mathcal{R}}_{\mathrm{d},m'}^{\mathrm{s}}>\mathcal{R} _{\mathrm{d},m^{\prime}}^{\mathrm{t}} \right) , \\
\mathcal{P} _{\mathrm{d},m'}^{\mathrm{s}}&=\mathrm{Pr}\left( \overline{\mathcal{R}}_{\mathrm{d},m'}^{\mathrm{s}}<\mathcal{R} _{\mathrm{d},m^{\prime}}^{\mathrm{t}} \right) ,
\end{align}
where $\mathcal{R} _{\mathrm{d},m}^{\mathrm{t}}$ and $\mathcal{R} _{\mathrm{d},m'}^{\mathrm{t}}$ denote the target rates of UT $m$ and UT $m'$, respectively. The following theorem provides a pair of lower bounds for the OPs of $m$ and $m'$, along with exact closed-form expressions in the case where $N=M$.
\begin{theorem} \label{SC_OP}
The lower bounds of $\mathcal{P} _{\mathrm{d},m}^{\mathrm{s}}$ and $\mathcal{P} _{\mathrm{d},m'}^{\mathrm{s}}$ satisfy
\begin{subequations}
\begin{align}
\!&\!\tilde{\mathcal{P}} _{\mathrm{d},m}^{\mathrm{s}}\!=\!F_{\lambda}\!\left(\frac{\varrho _m}{s_{m}^{\star}}\right) \!\!+\!F_{\lambda}\!\left( \frac{\varrho _{m'}}{s_{m}^{\star}} \right)\!\!-\!F_{\lambda}\!\left( \frac{\varrho _m}{s_{m}^{\star}} \right)\! F_{\lambda}\!\left( \frac{\varrho _{m'}}{s_{m}^{\star}} \right)\! ,\\
\!&\!\tilde{\mathcal{P}} _{\mathrm{d},m'}^{\mathrm{s}}\!=\!F_{\lambda}\left({\varrho _{m'}}/{s_{m}^{\star}} \right),
\end{align}
\end{subequations}
respectively, where $\varrho_m=\frac{2^{\mathcal{R} _{\mathrm{d},m}^{\mathrm{t}}}\!-1}{\alpha _m\eta _{m}^{-1}}$, $\varrho_{m'}=\frac{2^{\mathcal{R} _{\mathrm{d},m^{\prime}}^{\mathrm{t}}}-1}{\eta _{m^{\prime}}^{-1}-2^{\mathcal{R} _{\mathrm{d},m^{\prime}}^{\mathrm{t}}}\alpha _m\eta _{m^{\prime}}^{-1}} $, and $F_{\lambda}\left( \cdot \right) $ denotes the cumulative distribution function (CDF) of the largest eigenvalue of the complex Wishart matrix given in \cite[Eq. (9)]{wishart}. When $N=M$, we can derive the exact expressions of the OPs as follows:
\begin{subequations}
\begin{align}
\mathcal{P} _{\mathrm{d},m}^{\mathrm{s}}&=1-\exp \left( -\frac{\varrho _m+\varrho _{m'}}{s_{m}^{\star}}\right),\\
\mathcal{P} _{\mathrm{d},m'}^{\mathrm{s}}&=1-\exp \left( -\frac{\varrho _{m'}}{s_{m}^{\star}} \right) 
\end{align}    
\end{subequations}
\end{theorem}
\begin{IEEEproof}
Please refer to Appendix~\ref{Proof_SC_OP_theorem}.
\end{IEEEproof}
\begin{corollary}\label{SC_OP_asy}
When $p\rightarrow \infty $, the lower bounds of the OPs under the S-C design satisfy
\begin{subequations}
 \begin{align}
\tilde{\mathcal{P}}_{\mathrm{d},m}^{\mathrm{s},\infty}&\approx \left( {M}/{p} \right) ^{MN}\left( \varrho _m ^{MN}+\varrho _{m'} ^{MN} \right) ,\\
\tilde{\mathcal{P}}_{\mathrm{d},m'}^{\mathrm{s},\infty}&\approx \left( {M\varrho _{m'}}/{p} \right) ^{MN} ,
\end{align}   
\end{subequations}
For the case of $N=M$, the asymptotic OPs of UT $m$ and UT $m'$ are given by
\begin{align}
\mathcal{P} _{\mathrm{d},m}^{\mathrm{s},\infty}\approx \frac{M\left( \varrho _m +\varrho _{m'} \right)}{p} , \quad
\mathcal{P} _{\mathrm{d},m^{\prime}}^{\mathrm{s},\infty}\approx \frac{M\varrho _{m'}}{p} .
\end{align}       
\end{corollary}
\begin{IEEEproof}
When $p\rightarrow \infty$, we have $s_{m}^{\star}\approx \frac{p}{M}$. By utilizing ${\lim}_{x\rightarrow 0}F_{\lambda}\left( x \right) = x^{MN}$ \cite[Eq. (16)]{wishart_cdf} and ${\lim}_{x\rightarrow 0}\exp \left( -x \right) =1-x$ \cite[Eq. (1.211.1)]{integral}, the results can be obtained.
\end{IEEEproof}
To provide additional insights for system design, we next perform analysis for the diversity order by utilizing the derived asymptotic OPs. Taking UT $m$ as an example, its diversity order under the S-C design is defined as follows: \cite{mimo}
\begin{align}
\mathcal{D} _{\mathrm{d},m}^{\mathrm{s}}=-\underset{p\rightarrow \infty}{\lim}\frac{\log _2\left( \mathcal{P} _{\mathrm{d},m}^{\mathrm{s},\infty}\left( p \right) \right)}{\log _2p}.
\end{align}
The diversity order determines how fast the OP decays with the transmit power, at high SNR, shown in a log-log scale.
\begin{remark}\label{downlink_SC_OP_remark}
Based on the results in \textbf{Corollary~\ref{SC_OP_asy}}, under the S-C design, the lower bounds for the OPs of UT $m$ and UT $m'$ exhibit a diversity order of $MN$. Additionally, when $N=M$, the exact OPs have a diversity order of one, indicating there is no error floor.
\end{remark}
\subsection{Communications-Centric Design}
In this subsection, we investigate the C-C design.
\subsubsection{Performance of Communications}\label{Section: Downlink C_C Performance of Communications}
Since $\mathbf{H}_{\mathrm{d},i}$ has the same statistical properties as $\mathbf{H}_{\mathrm{d},i} \mathbf{U}$ for $i\in \left\{ m,m^{\prime} \right\}$, we can leverage the findings from Theorem~\ref{Sensing_Rate_S_C_Theorem} to design the C-C precoding marix as $\mathbf{P}=\mathbf{U}\bm{\Delta} _{\mathrm{c}}^{\frac{1}{2}}$, where $\bm{\Delta} _{\mathrm{c}}=\mathsf{diag}\left\{ c_1, \ldots,c_M \right\} $ satisfies $\sum_{m=1}^{M}c_{m}\leq p$ and $c_m\geq0$ for $m=1,\ldots,M$. Hence, the optimal precoding matrix is given by $\mathbf{P}_{\rm{c}}={\mathbf{U}}(\mathsf{diag}\left\{ c_1^{\star}, \ldots,c_M^{\star} \right\})^{\frac{1}{2}} $, where the optimal power allocation policy $\{c_{m}^{\star}\}_{m=1}^{M}$ is determined from the problem as follows:
\begin{align}\label{CC_optimal}
\underset{c_m}{\min}\,-\sum\nolimits_{m=1}^M&{\log _2\left( 1+\frac{\alpha _m\eta _{m}^{-1}c_m}{\left\| \mathbf{v}_{\mathrm{d},m} \right\|^2} \right) }\notag\\
&-\log _2\left( 1+\frac{\alpha _{m^{\prime}}\eta _{m^{\prime}}^{-1}c_m}{\alpha _m\eta _{m^{\prime}}^{-1}c_m+\left\| \mathbf{v}_{\mathrm{d},m} \right\|^2} \right),\notag
\\
\mathrm{s}.\mathrm{t}.\ \sum\nolimits_{m=1}^M&{c_m}-p=0, \quad c_m\geqslant 0,
\end{align}
which is convex and can be solved by using Karush–Kuhn–Tucker (KKT) conditions. The maximum sum CR achieved by $\mathbf{P}_{\rm{c}}$ is expressed as follows:
\begin{align}
\overline{\mathcal{R}}_{\mathrm{d},\mathrm{c}}^{\mathrm{c}}=\sum\nolimits_{m=1}^M(\overline{\mathcal{R}}_{\mathrm{d},m}^{\mathrm{c}}+ \overline{\mathcal{R}}_{\mathrm{d},m'}^{\mathrm{c}}),
\end{align}  
where $\overline{\mathcal{R}}_{\mathrm{d},m}^{\mathrm{c}}=\log _2\left( 1+\frac{\alpha _m\eta _{m}^{-1}c_{m}^{\star}}{\left\| \mathbf{v}_{\mathrm{d},m} \right\|^2} \right)$ and $\overline{\mathcal{R}}_{\mathrm{d},m'}^{\mathrm{c}}=\log _2\left( 1
+\frac{\alpha _{m^{\prime}}\eta _{m^{\prime}}^{-1}c_{m}^{\star}}{\alpha _m\eta _{m^{\prime}}^{-1}c_{m}^{\star}+\left\| \mathbf{v}_{\mathrm{d},m'} \right\|^2} \right)$ denote the CR of UT $m$ and UT $m'$ achieved by the C-C design, respectively. 

It is challenging to get a closed-form expression of the sum ECR $\mathcal{R} _{\mathrm{d},\mathrm{c}}^{\mathrm{c}}=\mathbb{E} \{ \overline{\mathcal{R}}_{\mathrm{d},\mathrm{c}}^{\mathrm{c}} \} $. Hence, to get more insights, we study its high-SNR performance in the following lemma.
\begin{lemma}
As $p\rightarrow \infty $, under the C-C design, an upper bound for the sum ECR is given by 
\begin{align} \label{CC_CR_asy_upper}
\tilde{\mathcal{R}}_{\mathrm{d},\mathrm{c}}^{\mathrm{c},\infty}\approx &M\left( \log _2p-\log _2M+\int_0^{\infty}{\log _2xf_{\gamma}\left( x \right) dx} \right) \notag\\
&+\sum\nolimits_{m=1}^M{\log _2\eta _{m}^{-1}}.
\end{align}
In the scenario where $N=M$, the asymptotic sum ECR in the high SNR-regime is derived as
\begin{align}\label{CC_CR_asy}
\mathcal{R} _{\mathrm{d},\mathrm{c}}^{\mathrm{c},\infty}\!\approx \!M\!\left(\! \log _2p\!-\!\log _2M\!-\!\frac{\mathcal{C}}{\ln 2} \right) \!+\!\sum\nolimits_{m=1}^M{\!\log _2\eta _{m}^{-1}}.
\end{align}
\end{lemma}
\begin{IEEEproof}
It is readily shown that the in the high-SNR regime, $c_m^{\star}$ satisfies $c_m^{\star}=\frac{p}{M}$ for $m=1,\ldots,M$. On this basis, the final results follow immediately by using similar steps in obtaining Corollary~\ref{SC_ECR_infinity}.
\end{IEEEproof}
\begin{remark} \label{CC_slope}
The results in \eqref{CC_CR_asy_upper} and \eqref{CC_CR_asy} suggest that the high-SNR slope of the sum ECR under the C-C design is $M$.
\end{remark}
Now we consider the OPs of UT $m$ and UT $m'$, which are defined as follows:
\begin{align}
\mathcal{P} _{\mathrm{d},m}^{\mathrm{c}}&=1-\mathrm{Pr}\left( \overline{\mathcal{R}}_{\mathrm{d},m}^{\mathrm{c}}>\mathcal{R} _{\mathrm{d},m}^{\mathrm{t}}, \overline{\mathcal{R}}_{\mathrm{d},m'}^{\mathrm{c}}>\mathcal{R} _{\mathrm{d},m^{\prime}}^{\mathrm{t}} \right) ,\\
\mathcal{P} _{\mathrm{d},m'}^{\mathrm{c}}&=\mathrm{Pr}\left( \overline{\mathcal{R}}_{\mathrm{d},m'}^{\mathrm{c}}<\mathcal{R} _{\mathrm{d},m^{\prime}}^{\mathrm{t}} \right).
\end{align}
\begin{lemma} \label{CC_OP_asy}
When $p\rightarrow \infty $, a pair of lower bounds for the OPs of UT $m$ and UT $m'$ under the C-C design yields
\begin{subequations}
\begin{align}
\tilde{\mathcal{P}}_{\mathrm{d},m}^{\mathrm{c},\infty}&\approx \left( {M}/{p} \right) ^{MN}\left( \varrho _m ^{MN}+\varrho _{m'} ^{MN} \right) ,\\
\tilde{\mathcal{P}}_{\mathrm{d},m'}^{\mathrm{c},\infty}&\approx \left( {M\varrho _{m'}}/{p} \right) ^{MN} ,
\end{align}    
\end{subequations}
For the case of $N=M$, the asymptotic OPs of $m$ and $m'$ in the high-SNR regime are given by
\begin{align}
\mathcal{P} _{\mathrm{d},m}^{\mathrm{c},\infty}\approx \frac{M\left( \varrho _m +\varrho _{m'} \right)}{p} , \quad
\mathcal{P} _{\mathrm{d},m^{\prime}}^{\mathrm{c},\infty}\approx \frac{M\varrho _{m'}}{p} .
\end{align} 
\end{lemma}
\begin{IEEEproof}
Similar to the proof of Corollary~\ref{SC_OP_asy}.
\end{IEEEproof}
\vspace{-5pt}
\begin{remark}\label{downlink_CC_OP_remark}
According to Lemma~\ref{CC_OP_asy}, the lower bounds for the OPs of UT $m$ and UT $m'$ demonstrate a diversity order of $MN$ under the C-C design. Furthermore, when $N=M$, the exact OPs exhibit a diversity order of one.
\end{remark}
By combining the results in Sections \ref{Section: Downlink S_C Performance of Communications} and \ref{Section: Downlink C_C Performance of Communications}, we find that in the high-SNR regime, the power allocation matrices satisfy $\lim_{p\rightarrow}\bm{\Delta} _{\mathrm{s}}\approx\lim_{p\rightarrow}\bm{\Delta} _{\mathrm{c}}\approx\frac{p}{M}{\mathbf{I}}_M$. On this basis, we have the following corollary.
\begin{corollary}\label{Corollary_Compare_CR_Downlink}
In the high-SNR regime, the CRs achieved by the S-C and C-C designs satisfy 
\begin{align}
&\overline{\mathcal{R} }_{\mathrm{d},m}^{\mathrm{s}}\approx\overline{\mathcal{R} }_{\mathrm{d},m}^{\mathrm{c}}\approx\log_2(1+{\alpha _m\eta _{m}^{-1}pM^{-1}}{\left\| \mathbf{v}_{\mathrm{d},m} \right\|^{-2}}),\\
&\overline{\mathcal{R} }_{\mathrm{d},m'}^{\mathrm{s}}\approx\overline{\mathcal{R} }_{\mathrm{d},m'}^{\mathrm{c}}\approx\log_2\left(1+\frac{\alpha _{m^{\prime}}\eta _{m^{\prime}}^{-1}p}{\alpha _m\eta _{m'}^{-1}p+M\left\| \mathbf{v}_{\mathrm{d},m'} \right\|^2}\right).
\end{align}
\end{corollary}
\vspace{-5pt}
\begin{IEEEproof}
This corollary can be directly proved by using the fact that ${\lim}_{x\rightarrow \infty}\log _2\left( 1+x \right) \approx \log _2x$.
\end{IEEEproof}
\vspace{-5pt}
\begin{remark} \label{CR_compare}
The results in \textbf{Corollary} \ref{Corollary_Compare_CR_Downlink} suggest that the CR achieved by the C-C design has the same asymptotic behaviour as that achieved by the S-C design, which means that they have the same high-SNR slope and diversity order.
\end{remark}
\subsubsection{Performance of Sensing}
When the precoding matrix $\mathbf{P}_{\mathrm{c}}$ is used, the SR is expressed as follows:
\begin{align} \label{CC_SR_exact}
\overline{\mathcal{R}} _{\mathrm{d},\mathrm{s}}^{\mathrm{c}}=ML^{-1}\sum\nolimits_{m=1}^M{\log _2\left( 1+L\lambda _m c_{m}^{\star} \right)}.
\end{align}
To account for the statistics of $c_{m}^{\star}$, we define the average SR as $\mathcal{R}_{\mathrm{d},\mathrm{s}}^{\mathrm{c}}=\mathbb{E} \{ \overline{\mathcal{R}}_{\mathrm{d},\mathrm{s}}^{\mathrm{c}} \} $ to numerically evaluate the average sensing performance. The high-SNR behaviour of $\mathcal{R}_{\mathrm{d},\mathrm{s}}^{\mathrm{c}}$ is described by the following theorem.
\begin{theorem}
When $p\rightarrow\infty$, the asymptotic SR under the C-C design satisfies
\begin{align}\label{CC_SR_Asymptotic}
\mathcal{R} _{\mathrm{d},\mathrm{s}}^{\mathrm{c},\infty}
\approx\frac{M^2}{L}\left(\log_2{p}+\frac{1}{M}\sum\nolimits_{m=1}^{M}\log_2\left(\frac{L\lambda_m}{M}\right)\right).
\end{align}
\end{theorem}
\begin{IEEEproof}
Similar to the proof of Theorem~\ref{Sensing_Rate_S_C_Theorem}.
\end{IEEEproof}

\begin{remark} \label{SR_compare}
The results in \eqref{CC_SR_Asymptotic} indicates that the high-SNR slope of the SR achieved by the C-C design is $\frac{M^2}{L}$. Moreover, by comparing \eqref{Sensing_Rate_S_C_Asymptotic} with \eqref{CC_SR_Asymptotic}, we find that the SR achieved by the C-C design involves the same asymptotic behaviour as that achieved by the S-C design.
\end{remark}
\vspace{-6pt}
\begin{remark} \label{Sc_CC_compare}
Based on the findings in \textbf{Remarks \ref{CR_compare}} and \textbf{\ref{SR_compare}}, we find that 1) the C-C design degenerates to the S-C design in the high-SNR regime and vice versa; and 2) both CR and SR increase linearly with the BS antenna number $M$. 
\end{remark}
\vspace{-7pt}
\subsection{Pareto Optimal Design}
In real-world scenarios, the precoding matrix $\mathbf{P}$ can be tailored to meet various quality of service requirements, leading to a tradeoff between communication and sensing performance in the downlink scenario. To assess this tradeoff, we use the Pareto boundary of the SR-CR region. The Pareto boundary is composed of SR-CR pairs where it is not possible to enhance one of the two rates without concomitantly diminishing the other \cite{pareto}. Specifically, we design the precoding matrix as $\mathbf{P}={\mathbf U}{\mathsf{diag}}\left\{p_1,\ldots,p_M\right\}$ with $\sum_{m=1}^{M}p_m\leq p$ and $p_m\geq0$. Accordingly, any rate-tuple on the Pareto boundary of the downlink rate region can be obtained via solving the following optimization problem: \cite{pareto}
\begin{equation}
\begin{split}\label{Problem_CR_SR_Tradeoff}
\max_{\mathbf{p},\mathcal{R}_{\rm{d}}}\ \ \mathcal{R}_{\rm{d}}, \ \ \mathrm{s}.\mathrm{t}.\ \ &\mathcal{R} _{\rm{d},\mathrm{s}}\ge \rho \mathcal{R}_{\rm{d}} ,\,\mathcal{R} _{\rm{d},\mathrm{c}}\ge \left( 1-\rho \right)\mathcal{R}_{\rm{d}},\\
&\sum\nolimits_{m=1}^{M}p_m\le p,\,p_m\ge 0,
\end{split}
\end{equation}
where $\rho\in\left[0,1\right]$ is a particular rate-profile parameter. Problem \eqref{Problem_CR_SR_Tradeoff} is convex and can be solved by using standard convex problem solvers, e.g., CVX. For a given value of $\rho$, let $\mathcal{R} _{\rm{d},\mathrm{s}}^{\rho}$ and $\mathcal{R} _{\rm{d},\mathrm{c}}^\rho$ represent the average SR and the sum ECR achieved by the corresponding optimal precoding matrix, respectively. Hence, we have $\mathcal{R} _{\mathrm{d},\mathrm{s}}^{\rho}\in [ \mathcal{R} _{\mathrm{d},\mathrm{s}}^{\mathrm{c}},\mathcal{R} _{\mathrm{d},\mathrm{s}}^{\mathrm{s}} ] $ and $\mathcal{R} _{\rm{d},\mathrm{c}}^\rho \in [ \mathcal{R} _{\rm{d},\mathrm{c}}^{\rm{s}} ,  \mathcal{R} _{\rm{d},\mathrm{c}}^{\rm{c}}]$ with $\mathcal{R} _{\rm{d},\mathrm{s}}^{1}=\mathcal{R} _{\rm{d},\mathrm{s}}^{\rm{s}}$ and $\mathcal{R} _{\rm{d},\mathrm{c}}^{0}=\mathcal{R} _{\rm{d},\mathrm{c}}^{\rm{c}}$, which yields the following corollary.
\begin{table}[!t]
\center
\begin{tabular}{|c|c|c|c|}\hline
\multicolumn{1}{|c|}{\multirow{2}{*}{System}} & \multicolumn{2}{c|}{CR} & SR \\ \cline{2-4}
 & $\mathcal{D}$  & $\mathcal{S}$ & $\mathcal{S}$ \\ \hline
 ISAC (S-C) & $1$ & $M$  & $M^2/L$  \\ \hline
  ISAC (C-C) & $1$ & $M$  & $M^2/L$  \\ \hline
 ISAC (Pareto Optimal) & $1$ & $M$  & $M^2/L$  \\ \hline
 FDSAC & $1$ & $\kappa M$  & $\left(1-\kappa\right)M^2/L$  \\ \hline
\end{tabular}
\caption{Downlink Diversity Order ($\mathcal{D}$) \\and High-SNR Slope ($\mathcal{S}$)}
\vspace{-10pt}
\label{table1}
\end{table}
\vspace{-2pt}
\begin{corollary}\label{Corollary_SR_CR_Asymptotic_Downlink}
In the high-SNR regime, the asymptotic behavior of $(\mathcal{R} _{\mathrm{d},\mathrm{s}}^{\rho},\mathcal{R} _{\rm{d},\mathrm{c}}^\rho)$ is the same as $(\mathcal{R} _{\mathrm{d},\mathrm{s}}^{\mathrm{c}},\mathcal{R} _{\rm{d},\mathrm{c}}^{\rm{c}})$ and $(\mathcal{R} _{\mathrm{d},\mathrm{s}}^{\mathrm{s}},\mathcal{R} _{\rm{d},\mathrm{c}}^{\rm{s}})$.
\end{corollary}
\begin{IEEEproof}
This corollary can be proved by applying the results in \textbf{Remark~\ref{Sc_CC_compare}} as well as the Sandwich theorem.
\end{IEEEproof}
\vspace{-3pt}
\begin{remark}
The results in \textbf{Corollary \ref{Corollary_SR_CR_Asymptotic_Downlink}} suggest that any SR-CR tuple on the Pareto boundary has the same asymptotic behaviour in the high-SNR regime.
\end{remark}

\vspace{-12pt}
\subsection{Performance of FDSAC}
The baseline scenario we are examining is FDSAC, which involves dividing the overall bandwidth into two sub-bands: one exclusively for sensing and the other for communications. Additionally, the total power is also partitioned into two portions for S\&C, respectively. Specifically, we assume $\kappa$ fraction of the total bandwidth and $\mu $ fraction of the total power is used for communications. In this case, the SR and the sum ECR are given by
{\setlength{\abovedisplayskip}{3pt} 
\setlength{\belowdisplayskip}{1pt}
\begin{align}
\mathcal{R} _{\mathrm{d},\mathrm{s}}^{\mathrm{f}}\!&=\!\frac{M(1\!-\!\kappa )}{L}\!\max _{\sum_{m=1}^M{a_m}\le (1-\mu )p}\sum\nolimits_{m=1}^M\!{\log _2\!\left( \!1\!+\!\frac{L\lambda _m}{1\!-\!\kappa}a_m \!\right)},\\
\mathcal{R} _{\mathrm{d},\mathrm{c}}^{\mathrm{f}}\!&=\!\mathbb{E} \left\{ \max _{\sum_{m=1}^M{b_m}\le \mu p}\!\sum\nolimits_{m=1}^M{\!\kappa \log _2\!\left(\! 1\!+\!\frac{\alpha _m\eta _{m}^{-1}b_m}{\kappa\left\| \mathbf{v}_{\mathrm{d},m} \right\|^2}\! \right)} \right. \nonumber\\
&~~~~~~~\left. +\kappa \log _2\!\left( \!1\!+\!\frac{\alpha _{m^{\prime}}\eta _{m^{\prime}}^{-1}b_m}{\alpha _m\eta _{m^{\prime}}^{-1}b_m\!+\!\kappa\left\| \mathbf{v}_{\mathrm{d},m'} \right\|^2} \!\right)\! \right\},
\end{align}}
respectively. It is worth noting that $(\mathcal{R} _{\mathrm{d},\mathrm{s}}^{\mathrm{f}},\mathcal{R} _{\mathrm{d},\mathrm{c}}^{\mathrm{f}})$ can be discussed in the way we discuss $(\mathcal{R} _{\mathrm{d},\mathrm{s}}^{\mathrm{s}},\mathcal{R}_{\mathrm{d},\mathrm{c}}^{\mathrm{c}})$. 

Upon concluding all the analyses of downlink case, we consolidate the outcomes pertaining to diversity order and high-SNR slope within Table~\ref{table1}.

\begin{remark} \label{FDSAC_compare}
The results in Table~\ref{table1} suggest that FDSAC achieves the same diversity order as ISAC, while the high-SNR slopes of both SR and CR achieved by FDSAC are smaller than those achieved by ISAC. This fact means that both the SR and CR of ISAC increase faster with the SNR compared to FDSAC, i.e., ISAC is able to provide more DoFs than FDSAC in terms of both S\&C.  
\end{remark}

\subsection{Rate Region Characterization}
We now characterize the SR-CR region achieved by the downlink ISAC and FDSAC systems. In particular, let ${\mathcal{R}}_{\rm{d},\rm{s}}$ and ${\mathcal{R}}_{\rm{d},\rm{c}}$ denote the achievable SR and CR, respectively. Then, the rate regions achieved by ISAC and FDSAC are, respectively, given by
\begin{align}
\mathcal{C} _{\mathrm{d},\mathrm{i}}&=\!\left\{ ( \mathcal{R} _{\mathrm{d},\mathrm{s}},\mathcal{R} _{\mathrm{d},\mathrm{c}} ) \left|\!\! \begin{array}{c}
	\mathcal{R} _{\mathrm{d},\mathrm{s}}\!\in\!\! [ 0,\mathcal{R} _{\mathrm{d},\mathrm{s}}^{\rho} ] ,\mathcal{R} _{\mathrm{d},\mathrm{c}}\!\in\! [ 0,\mathcal{R} _{\mathrm{d},\mathrm{c}}^{\rho} ] ,\\
	\rho \in \left[ 0,1 \right]\\
\end{array} \right. \!\!\right\} ,\label{Rate_Regio_ISAC}\\
\mathcal{C} _{\mathrm{d},\mathrm{f}}&=\!\left\{ ( \mathcal{R} _{\mathrm{d},\mathrm{s}},\mathcal{R} _{\mathrm{d},\mathrm{c}} ) \left|\!\! \begin{array}{c}
	\mathcal{R} _{\mathrm{d},\mathrm{s}}\!\in \![ 0,\mathcal{R} _{\mathrm{d},\mathrm{s}}^{\mathrm{f}} ] ,\mathcal{R} _{\mathrm{d},\mathrm{c}}\!\in\! [ 0,\mathcal{R} _{\mathrm{d},\mathrm{c}}^{\mathrm{f}} ] ,\\
	\kappa \in [ 0,1 ] ,\mu \in [ 0,1 ]\\
\end{array} \right. \!\!\right\}  .\label{Rate_Regio_FDSAC}
\end{align}  
\begin{theorem}\label{rate_region}
The above rate regions satisfy $\mathcal{C} _{\rm{d},\mathrm{f}}\subseteq \mathcal{C} _{\rm{d},\mathrm{i}}$. 
\end{theorem}
\begin{IEEEproof}
Please refer to Appendix~\ref{Proof_rate_region}.
\end{IEEEproof}
According to Theorem~\ref{rate_region}, the downlink rate region achieved by ISAC entirely encompasses that achieved by FDSAC. This is mainly attributed to ISAC’s integrated exploitation of both spectrum and power resources.

\section{Uplink Performance}\label{uplink}
In this section, we analyze the performance of uplink ISAC under the S-C and C-C designs according to the interference cancellation order of the two-stage SIC. Also, FDSAC is considered as the baseline and the uplink rate regions are provided.
\subsection{Sensing-Centric Design}
In the S-C design, the BS first decodes the communication signal $\{\mathbf{s}_{\mathrm{u},m},\mathbf{s}_{\mathrm{u},m'}\}_{m=1}^{M}$ transmitted by UTs, considering the sensing signal $\mathbf{GX}_{\rm{s}}$ as interference. Then, the BS subtracts the decoded communication signal from the received signal, utilizing the remaining part for sensing the target response matrix $\mathbf{G}$.

\subsubsection{Performance of Communications}
From a worst-case design standpoint, the aggregate interference-plus-noise at $l$th time slot ${\bm\xi}  _{m,l}=\mathbf{Gx}_{{\rm{s}},l}+\mathbf{n}_{{\mathrm{u}},l}$ is regarded as the Gaussian noise \cite{GaussianNoise}. On this basis, we conclude the following lemma.
\begin{lemma}\label{uplink_noise_2}
At the $l$th time slot, the uplink SNR of UT $m$ and SINR of UT $m'$ are, respectively, given by 
\begin{align}
\gamma _{\mathrm{u},m,l}^{\mathrm{s}}&=\frac{\alpha _m\eta _{m}^{-1}p_{\mathrm{c}}}{\left( \mathbf{Q}^{-\mathsf{H}}\mathbf{Q}^{-1} \right) _{m,m}\sigma _{\mathrm{s},l}^{2}}, \\
\gamma _{\mathrm{u},m^{\prime},l}^{\mathrm{s}}&=\frac{\alpha _{m'}\eta _{m'}^{-1}p_{\mathrm{c}}}{\alpha _m\eta _{m}^{-1}p_{\mathrm{c}}+\left( \mathbf{Q}^{-\mathsf{H}}\mathbf{Q}^{-1} \right) _{m,m}\sigma _{\mathrm{s},l}^{2}},
\end{align}
where $\sigma _{\mathrm{s},l}^{2}=| \mathbf{x}_{\mathrm{s},l}^{\mathsf{H}}\mathbf{Rx}_{\mathrm{s},l} |+1$. 
\end{lemma}
\begin{IEEEproof}
Utilizing the properties $\mathbf{g}_n\sim{\mathcal{CN}}\left({\mathbf{0}},\mathbf{R}\right)$ for $n=1,\ldots,N$ and $\mathbb{E} \left\{ \mathbf{g}_n\mathbf{g}_{n'}^{\mathsf{H}} \right\} =\mathbf{0}$ for $n\ne n'$, as well as $\mathbf{n}_{\mathrm{u},l}\sim{\mathcal{CN}}\left({\mathbf{0}},\mathbf{I}_M\right)$, we derive the mean and variance of ${\bm{\xi}}_{m,l}$ as follows:
\begin{subequations}
\begin{align}
\mathbb{E} \left\{ {\bm\xi}_{m,l} \right\}&=\mathbf{0},\\
\mathbb{E} \left\{ {\bm\xi}_{m,l}{\bm\xi}_{m,l}^{\mathsf{H}} \right\} &=\mathbb{E} \left\{ \mathbf{Gx}_{\mathrm{s},l}\mathbf{x}_{\mathrm{s},l}^{\mathsf{H}}\mathbf{G}^{\mathsf{H}} \right\} +\mathbf{I}_M\notag\\
&=\left| \mathbf{x}_{\mathrm{s},l}^{\mathsf{H}}\mathbf{Rx}_{\mathrm{s},l} \right|\mathbf{I}_M+\mathbf{I}_M\triangleq\sigma _{\mathrm{s},l}^{2}\mathbf{I}_M.
\end{align}    
\end{subequations}
Therefore, for each UT pair, when treating the aggregate interference-plus-noise at the $l$th time slot as Gaussian noise, each of its elements exhibits a zero mean and a variance of $\sigma_{{\rm{s}},l}^{2}$. Since the detection vector $\mathbf{v}_{\mathrm{u},m}$ is normalized, the SNR and the SINR can be derived straightforwardly.
\end{IEEEproof}
\begin{theorem}\label{uplink_SC_CR}
The uplink sum ECR under the S-C design is given by
\begin{align}\label{uplink_SC_ECR}
\mathcal{R} _{\mathrm{u},\mathrm{c},l}^{\mathrm{s}}=-\frac{1}{\ln 2}\sum\nolimits_{m=1}^M{\exp \left( \frac{\sigma _{\mathrm{s},l}^{2}}{p_{\mathrm{c}}\delta _m} \right) \mathrm{Ei}\left( -\frac{\sigma _{\mathrm{s},l}^{2}}{p_{\mathrm{c}}\delta _m} \right)},
\end{align}
where $\delta _m=\alpha _m\eta _{m}^{-1}+\alpha _{m^{\prime}}\eta _{m^{\prime}}^{-1}$. When $p_{\rm{c}}\rightarrow \infty $, the asymptotic ECR is expressed as follows:
\begin{align}\label{uplink_SC_ECR_asy}
\mathcal{R} _{\mathrm{u},\mathrm{c},l}^{\mathrm{s},\infty}\approx M\left( \log _2p_{\mathrm{c}}+\log _2\delta _m-\log _2\sigma _{\mathrm{s},l}^{2}-\mathcal{C}/\ln{2} \right) .    
\end{align}
\end{theorem}
\begin{IEEEproof}
Please refer to Appendix~\ref{Proof_uplink_CC_ECR}.
\end{IEEEproof}
\begin{remark}
The results in \eqref{uplink_SC_ECR_asy} indicate that the high-SNR slope of the uplink sum ECR achieved by the S-C design is given by $M$.
\end{remark}
The following theorem provides the closed-form expressions for the OP as well as its high-SNR approximations.
\begin{theorem}\label{uplink_SC_OP}
Under the uplink S-C design, the OP for the $m$th UT pair is given by
\begin{align}
\mathcal{P} _{\mathrm{u},m,l}^{\mathrm{s}}=1-\exp \left( -\sigma _{\mathrm{s},l}^{2}\frac{2^{\mathcal{R} _{\mathrm{u}}^{\mathrm{t}}}-1}{p_{\mathrm{c}}\delta _m} \right) .    
\end{align}
where $\mathcal{R} _{\mathrm{u}}^{\mathrm{t}}$ denotes the target rate for the $m$th pair. When $p_{\rm{c}}\rightarrow \infty $, its high-SNR approximation satisfies
\begin{align} \label{uplink_SC_OP_asy}
\mathcal{P} _{\mathrm{u},m,l}^{\mathrm{s},\infty}\approx \sigma _{\mathrm{s},l}^{2}\frac{2^{\mathcal{R} _{\mathrm{u}}^{\mathrm{t}}}-1}{p_{\mathrm{c}}\delta _m}.
\end{align}
\end{theorem}
\begin{IEEEproof}
The OP of the $m$th UT pair is calculated as 
\begin{align}
\mathcal{P} _{\mathrm{u},m,l}^{\mathrm{s}}&=\mathrm{Pr}\left( \frac{1}{\left( \mathbf{Q}^{-\mathsf{H}}\mathbf{Q}^{-1} \right) _{m,m}}<\sigma _{\mathrm{s},l}^{2}\frac{2^{\mathcal{R} _{\mathrm{u}}^{\mathrm{t}}}-1}{p_{\mathrm{c}}\delta _m} \right) \notag\\
&=F_Q\left( \sigma _{\mathrm{s},l}^{2}\frac{2^{\mathcal{R} _{\mathrm{u}}^{\mathrm{t}}}-1}{p_{\mathrm{c}}\delta _m} \right) , 
\end{align}
where $F_Q\left(x\right)$ denotes the CDF of $\frac{1}{\left( \mathbf{Q}^{-\mathsf{H}}\mathbf{Q}^{-1} \right) _{m,m}}$. Since $\frac{1}{\left( \mathbf{Q}^{-\mathsf{H}}\mathbf{Q}^{-1} \right) _{m,m}}$ is  exponentially distributed \cite{signalalignment,Ding_relay}, we have $F_Q\left( \sigma _{\mathrm{s},l}^{2}\frac{2^{\mathcal{R} _{\mathrm{u}}^{\mathrm{t}}}-1}{p_{\mathrm{c}}\delta _m} \right) =1-\exp \left(-\sigma _{\mathrm{s},l}^{2} \frac{2^{\mathcal{R} _{\mathrm{u}}^{\mathrm{t}}}-1}{p_{\mathrm{c}}\delta _m} \right) $. When $p_{\rm{c}}\rightarrow \infty $, by utilizing the property $\lim _{x\rightarrow 0}\exp \left( -x \right) =1-x$, we can obtain \eqref{uplink_SC_OP_asy}.
\end{IEEEproof}
\begin{remark}
According to the result in \eqref{uplink_SC_OP_asy}, the diversity order for the OP achieved by the uplink S-C design is given by one.
\end{remark}
\subsubsection{Performance of Sensing}
Removing the decoded communication signal from the received signal, we now focus on the rest part which is used for sensing. The SR is dependent on the sensing signal $\mathbf{X}_{\mathrm{s}}$ and the maximum achievable uplink SR by S-C design is expressed as
\begin{align}
\mathcal{R} _{\mathrm{u},\mathrm{s}}^{\mathrm{s}}\!=\!\max _{\mathsf{tr}\left( \mathbf{X}_{\mathrm{s}}\mathbf{X}_{\mathrm{s}}^{\mathsf{H}} \right) \le Lp_s}\!ML^{-1}\log _2\det \left( \mathbf{I}_M\!+\!\mathbf{X}_{\mathrm{s}}^{\mathsf{H}}\mathbf{RX}_{\mathrm{s}} \right) ,
\end{align}
where $p_{\rm{s}}$ represents the per-symbol power budget for the uplink sensing signal.
\begin{theorem}\label{Sensing_Rate_C_SIC_Theorem}
The maximum SR achieved under the S-C design is given by
\begin{align} 
\mathcal{R} _{\mathrm{u},\mathrm{s}}^{\mathrm{s}}=ML^{-1}\sum\nolimits_{m=1}^M{\log _2\left( 1+\lambda _m\zeta  _{m}^{\mathrm{s}} \right)},
\end{align}
where $\zeta  _{m}^{\mathrm{s}}=\max\{0,\frac{1}{\nu}-\frac{1}{\lambda_m}\}$ with $\sum_{m=1}^{M}\max\{0,\frac{1}{\nu}-\frac{1}{\lambda_m}\}=Lp_{\rm{s}}$. The maximum SR is achieved when $\mathbf{X}_{\mathrm{s}}\mathbf{X}_{\mathrm{s}}^{\mathsf{H}}={\mathbf{U}}{\bm\Delta}_{\varepsilon}^{\rm{s}}{\mathbf{U}}^{\mathsf{H}}$, where ${\bm\Delta}_{\varepsilon}^{\rm{s}}={\mathsf{diag}}\left\{\zeta  _{1}^{\mathrm{s}},\ldots,\zeta  _{M}^{\mathrm{s}}\right\}$.
When $p_{\rm{s}}\rightarrow\infty$, the asymptotic SR is derived as follows:
\begin{align}
\mathcal{R} _{\mathrm{u},\mathrm{s}}^{\mathrm{s},\infty}
\approx\frac{M^2}{L}\left(\log_2{p_{\rm{s}}}+\frac{1}{M}\sum\nolimits_{m=1}^{M}\log_2\left(\frac{L\lambda_m}{M}\right)\right).
\end{align}
\end{theorem}
\begin{IEEEproof}
Similar to the proof of Theorem~\ref{Sensing_Rate_S_C_Theorem}.
\end{IEEEproof}
\begin{remark}
The high-SNR slope of the uplink SR achieved under the S-C design is $\frac{M^2}{L}$.
\end{remark}
\subsection{Communications-Centric Design}
In the C-C design, the target response signal is initially estimated by treating the communication signal as interference, followed by decoding the communication signals of UT $m$ and UT $m'$ after removing the sensing signal.
\subsubsection{Performance of Sensing}
From a worst-case design standpoint, the aggregate interference-plus-noise $\mathbf{\Phi }=\sum\nolimits_{m=1}^M\left(\sqrt{\alpha _m\eta _{m}^{-1}p_{\rm{c}}}\mathbf{H}_{\mathrm{u},m}\mathbf{w}_{\mathrm{u},m}\mathbf{s}_{\mathrm{u},m}^{\mathsf{H}}+\right.$ $\left.\sqrt{\alpha _{m^{\prime}}\eta _{m^{\prime}}^{-1}p_{\rm{c}}}\mathbf{H}_{\mathrm{u},m'}\mathbf{w}_{\mathrm{u},m'}\mathbf{s}_{u,m^{\prime}}^{H}\right)+\mathbf{N}_{\mathrm{u}}$ can be treated as the Gaussian noise \cite{GaussianNoise}. The uplink SR under the C-C design is provided in the following lemma.
\begin{lemma}\label{uplink_noise_1}
The maximum achievable uplink SR by the C-C design is expressed as follows:
\begin{align}
\mathcal{R} _{\mathrm{u},\mathrm{s}}^{\mathrm{c}}\!=\!\max _{\mathsf{tr}\left( \mathbf{X}_{\mathrm{s}}\mathbf{X}_{\mathrm{s}}^{\mathsf{H}} \right) \le Lp_s}\!ML^{-1}\!\log _2\det\! \left( \mathbf{I}_M\!+\!\sigma _{\mathrm{c}}^{2}\mathbf{X}_{\mathrm{s}}^{\mathsf{H}}\mathbf{RX}_{\mathrm{s}} \right)  
\end{align}
with $\sigma _{\mathrm{c}}^{2}=p_{\mathrm{c}}\sum\nolimits_{m=1}^M{\phi _m}+1$ and ${\phi _m}=\alpha _m\eta _{m}^{-1}+\alpha _{m^{\prime}}\eta _{m^{\prime}}^{-1} $.
\end{lemma}
\begin{IEEEproof}
Please refer to Appendix~\ref{Proof_uplink_noise_1}.
\end{IEEEproof}
\begin{theorem}\label{Theorem_SR_ER}
The exact expression of the maximum SR under the C-C design is given by 
\begin{align} 
\mathcal{R} _{\mathrm{u},\mathrm{s}}^{\mathrm{c}}=ML^{-1}\sum\nolimits_{m=1}^M{\log _2\left( 1+\sigma _{\mathrm{c}}^{-2}\lambda _m\zeta_{m}^{\rm{c}} \right)},
\end{align}
where $\zeta  _{m}^{\mathrm{c}}=\max\{0,\frac{1}{\nu}-\frac{\sigma _{\mathrm{c}}^{2}}{\lambda_m}\}$ with $\sum_{m=1}^{M}\max\{0,\frac{1}{\nu}-\frac{\sigma _{\mathrm{c}}^{2}}{\lambda_m}\}=Lp_{\rm{s}}$. The maximum SR is achieved when $\mathbf{X}_{\mathrm{s}}\mathbf{X}_{\mathrm{s}}^{\mathsf{H}}={\mathbf{U}}{\bm\Delta}_{\varepsilon}^{\rm{c}}{\mathbf{U}}^{\mathsf{H}}$, where ${\bm\Delta}_{\varepsilon}^{\rm{c}}={\mathsf{diag}}\left\{\zeta  _{1}^{\mathrm{c}},\ldots,\zeta  _{M}^{\mathrm{c}}\right\}$.When $p_{\rm{s}}\rightarrow\infty$, the asymptotic SR satisfies
\begin{align}
\mathcal{R} _{\mathrm{u},\mathrm{s}}^{\mathrm{c},\infty}
\approx\frac{M^2}{L}\left(\log_2{p_{\rm{s}}}+\frac{1}{M}\sum\nolimits_{m=1}^{M}\log_2\left(\frac{L\lambda_m}{\sigma _{\mathrm{c}}^{2}M}\right)\right).
\end{align}
\end{theorem}
\begin{IEEEproof}
Similar to the proof of Theorem~\ref{Sensing_Rate_S_C_Theorem}.
\end{IEEEproof}
\begin{remark}
The high-SNR slope of the SR achieved under the uplink C-C design is $\frac{M^2}{L}$.
\end{remark}
\subsubsection{Performance of Communications}
After the target response is estimated, the sensing echo signal $\mathbf{G}\mathbf{X}$ can be subtracted from the received superposed S\&C signal. Therefore, under the C-C design, the SNR for UT $m$ and the SINR for UT $m'$ are, respectively, given by
\begin{align}
&\gamma _{\mathrm{u},m}^{\mathrm{c}}=\frac{\alpha _m\eta _{m}^{-1}p_{\mathrm{c}}}{\left( \mathbf{Q}^{-\mathsf{H}}\mathbf{Q}^{-1} \right) _{m,m}}, \\
&\gamma _{\mathrm{u},m^{\prime}}^{\mathrm{c}}=\frac{\alpha _{m'}\eta _{m'}^{-1}p_{\mathrm{c}}}{\alpha _m\eta _{m}^{-1}p_{\mathrm{c}}+\left( \mathbf{Q}^{-\mathsf{H}}\mathbf{Q}^{-1} \right) _{m,m}}.
\end{align} 
\begin{theorem}\label{uplink_CC_ECR}
The uplink sum ECR under the C-C design is given by
\begin{align}\label{uplink_CC_ECR_exa}
\mathcal{R} _{\mathrm{u},\mathrm{c}}^{\mathrm{c}}=-\frac{1}{\ln 2}\sum\nolimits_{m=1}^M{\exp \left( \frac{1}{p_{\mathrm{c}}\delta _m} \right) \mathrm{Ei}\left( -\frac{1}{p_{\mathrm{c}}\delta _m} \right)}.
\end{align} 
When $p_{\rm{c}}\rightarrow \infty $, the high-SNR approximation of the sum ECR is derived as follows:
\begin{align}\label{uplink_CC_ECR_asy}
 \mathcal{R} _{\mathrm{u},\mathrm{c}}^{\mathrm{c},\infty}\approx M\left( \log _2p_{\mathrm{c}}+\log _2\delta _m-\mathcal{C}/\ln{2} \right).
\end{align}
\end{theorem}
\begin{IEEEproof}
Similar to the proof of Theorem~\ref{uplink_SC_CR}.
\end{IEEEproof}
\begin{remark}
The results in \eqref{uplink_CC_ECR_asy} suggest that the high-SNR slope of the uplink sum ECR in the C-C design is $M$.
\end{remark}
We next investigate the uplink outage performance under the C-C design.
\begin{theorem}\label{uplink_CC_OP}
The closed-form expression of OP for the $m$th UT pair is given by 
{\setlength{\abovedisplayskip}{0pt} 
\setlength{\belowdisplayskip}{3pt}
\begin{align}\label{uplink_CC_OP_exa}
\mathcal{P} _{\mathrm{u},m}^{\mathrm{c}}=1-\exp \left( -\frac{2^{\mathcal{R} _{\mathrm{u}}^{\mathrm{t}}}-1}{p_{\mathrm{c}}\delta _m} \right) .   
\end{align}}
 When $p_{\rm{c}}\rightarrow \infty $, its high-SNR approximation satisfies 
 {\setlength{\abovedisplayskip}{4pt} 
\setlength{\belowdisplayskip}{4pt}
\begin{align}\label{uplink_CC_OP_asy}
\mathcal{P} _{\mathrm{u},m}^{\mathrm{c},\infty}\approx \frac{2^{\mathcal{R} _{\mathrm{u}}^{\mathrm{t}}}-1}{p_{\mathrm{c}}\delta _m}.
\end{align}}
\end{theorem}
\begin{IEEEproof}
Similar to the proof of Theorem~\ref{uplink_SC_OP}.
\end{IEEEproof}
\begin{remark}
According to the results in \eqref{uplink_CC_OP_asy}, a diversity of one is achievable for the OP of the $m$th UT pair under the C-C design. 
\end{remark}
\subsection{Discussion on SIC Ordering}\label{Section: Discussion on SIC Ordering}
Having characterized the CRs and SRs achieved by the C-C and S-C SIC orders, we now shift our focus on the influence of SIC ordering on the S\&C performance. 

Let us first consider the SR. By comparing \textbf{Theorem \ref{Sensing_Rate_C_SIC_Theorem}} with \textbf{Theorem \ref{Theorem_SR_ER}}, we obtain the following results.
\begin{remark}
The SRs achieved by the S-C and C-C designs have the same high-SNR slope given by $\frac{M^2}{L}$, which means that the SIC ordering has no influence on the high-SNR slope of the SR.
\end{remark}
\begin{corollary}\label{Corollary_SR_Power_Compare}
In the high-SNR regime, we can obtain $\lim_{p_{\rm{s}}\rightarrow\infty}(\mathcal{R} _{\mathrm{u},\mathrm{s}}^{\mathrm{s}}-\mathcal{R} _{\mathrm{u},\mathrm{s}}^{\mathrm{c}})\approx
\frac{M^2}{L}{\mathcal E}_{\rm s}$, where
\begin{align}
{\mathcal E}_{\rm s}=\log _2\sigma _{\mathrm{c}}^{2}=\log_2\left(p_{\mathrm{c}}\sum\nolimits_{m=1}^M{\phi _m}+1\right) >0.
\end{align}
\end{corollary}
\begin{remark}\label{Discussion_SIC_Sensing}
The above results suggest that the S-C design is superior to the C-C design in terms of the achievable SR by an SNR gap of ${\mathcal E}_{\rm s}$ in 3-dB units \cite{mimo}. In other words, to achieve the same SR as the S-C design, the C-C design has to consume more SNR of ${\mathcal E}_{\rm s}$ in 3-dB units to resist the IFI \cite{mimo}.
\end{remark}
Note that the SNR gap ${\mathcal E}_{\rm s}$ is a monotone increasing function of the communication power budget $p_{\rm c}$. This indicates that the SNR gap concerning the SR introduced by the SIC ordering is more highlighted for a higher communication SNR, which is as expected.

We next consider the sum-CR. By comparing the results in \textbf{Theorems \ref{uplink_SC_OP}} and \textbf{\ref{uplink_CC_OP}}, we find that the SIC ordering does not influence the diversity order of the sum-CR. We next compare \textbf{Theorem \ref{uplink_SC_CR}} with \textbf{Theorem \ref{uplink_CC_ECR}}, which leads to the following results.
\begin{remark}
The ECRs achieved by the S-C and C-C designs have the same high-SNR slope given by $M$, which means that the SIC ordering has no influence on the high-SNR slope of the ECR.
\end{remark}
\begin{corollary}\label{Corollary_CR_Power_Compare}
In the high-SNR regime, we can obtain $\lim_{p_{\rm{c}}\rightarrow\infty}(\mathcal{R} _{\mathrm{u},\mathrm{c}}^{\mathrm{c}}-\sum_{l=1}^{L}\mathcal{R} _{\mathrm{u},\mathrm{c},l}^{\mathrm{s}})\approx
M{\mathcal E}_{\rm c}$, where
\begin{align}
{\mathcal E}_{\rm c}&=\frac{1}{L}\sum\nolimits_{l=1}^{L}\log _2\sigma _{\mathrm{s},l}^{2}\notag\\
&=\frac{1}{L}\sum\nolimits_{l=1}^{L}\log_2\left(| \mathbf{x}_{\mathrm{s},l}^{\mathsf{H}}\mathbf{Rx}_{\mathrm{s},l} |+1\right) >0.
\end{align}
\end{corollary}
\begin{remark}\label{Discussion_SIC_Communications}
The above results suggest that the C-C design is superior to the S-C design in terms of the achievable ECR by an SNR gap of ${\mathcal E}_{\rm c}$ in 3-dB units \cite{mimo}. Or in other words, to achieve the same ECR as the C-C design, the S-C design has to consume more SNR of ${\mathcal E}_{\rm c}$ in 3-dB units to resist the interference from the sensing echo signal. Intuitively, this SNR gap ${\mathcal E}_{\rm c}$ increases with the sensing power budget $p_{\rm s}$.
\end{remark}

\begin{table}[!t]
\center
\begin{tabular}{|c|c|c|c|}\hline
\multicolumn{1}{|c|}{\multirow{2}{*}{System}} & \multicolumn{2}{c|}{CR} & SR \\ \cline{2-4}
 & $\mathcal{D}$  & $\mathcal{S}$ & $\mathcal{S}$ \\ \hline
 ISAC (S-C) & $1$ & $M$  & $M^2/L$  \\ \hline
  ISAC (C-C) & $1$ & $M$  & $M^2/L$  \\ \hline
 ISAC (Time-Sharing) & $1$ & $M$  & $M^2/L$  \\ \hline
 FDSAC & $1$ & $\kappa M$  & $\left(1-\kappa\right)M^2/L$  \\ \hline
\end{tabular}
\caption{Uplink Diversity Order ($\mathcal{D}$) \\and High-SNR Slope ($\mathcal{S}$)}
\label{table2}
\end{table}
\subsection{Rate Region Characterization}
We now characterize the uplink SR-CR region achieved by the NOMA-ISAC. By utilizing the time-sharing strategy \cite{mimo}, we apply the S-C design with probability $\tau$ and the C-C design with probability $1-\tau$. For a given $\tau$, the achievable rate tuple is represented by $\left( \mathcal{R} _{\mathrm{u},\mathrm{s}}^{\tau},\mathcal{R} _{\mathrm{u},\mathrm{c}}^{\tau} \right) $, where $\mathcal{R} _{\mathrm{u},\mathrm{s}}^{\tau}=\tau \mathcal{R} _{\mathrm{u},\mathrm{s}}^{\mathrm{s}}+\left( 1-\tau \right) \mathcal{R} _{\mathrm{u},\mathrm{s}}^{\mathrm{c}}$ and $\mathcal{R} _{\mathrm{u},\mathrm{c}}^{\tau}=\tau \mathcal{R} _{\mathrm{u},\mathrm{c}}^{\mathrm{s}}+\left( 1-\tau \right) \mathcal{R} _{\mathrm{u},\mathrm{c}}^{\mathrm{c}}$. Let $\mathcal{R} _{\mathrm{u},\mathrm{s}}$ and $\mathcal{R} _{\mathrm{u},\mathrm{c}}$ denote the achievable SR and CR, respectively. Then, the rate region achieved by uplink ISAC follows
\begin{align}
\mathcal{C} _{\mathrm{u},\mathrm{i}}\!=\!\left\{ \!\left( \mathcal{R} _{\mathrm{u},\mathrm{s}},\mathcal{R} _{\mathrm{u},\mathrm{c}} \right) \left| \!\!\begin{array}{c}
	\mathcal{R} _{\mathrm{u},\mathrm{s}}\!\in\! \left[ 0,\mathcal{R} _{\mathrm{u},\mathrm{s}}^{\tau} \right] ,\mathcal{R} _{\mathrm{u},\mathrm{c}}\!\in \!\left[ 0,\mathcal{R} _{\mathrm{u},\mathrm{c}}^{\tau} \right] ,\\
	\tau \in \left[ 0,1 \right]\\
\end{array} \right.\!\!\! \right\} .
\end{align}
\begin{remark}
Exploiting the Sandwich theorem, we obtain that any rate-tuple achieved by the time-sharing strategy yields the same high-SNR slopes and diversity order.
\end{remark}

\subsection{Performance of FDSAC}
We also consider uplink FDSAC as the baseline scenario, where a fraction $\kappa \in \left[0,1\right]$ of the total bandwidth is allocated for communications, while the remaining fraction $1-\kappa$ of the bandwidth is allocated for sensing. In this case, the sum ECR and the SR of FDSAC satisfy
\begin{align}
\mathcal{R} _{\mathrm{u},\mathrm{c}}^{\mathrm{f}}&=\!\sum\nolimits_{m=1}^M{\mathbb{E} \left\{ \kappa \log _2\left( 1\!+\!\frac{p_{\mathrm{c}}\delta _m}{\kappa \left( \mathbf{Q}^{-1}\mathbf{Q}^{-\mathsf{H}} \right) _{m,m}} \right) \right\}},\\
 \mathcal{R} _{\mathrm{u},\mathrm{s}}^{\mathrm{f}}&=\frac{\left( 1-\kappa \right) M}{L}\sum\nolimits_{m=1}^M{\log _2\left( 1+\frac{\lambda _m\zeta _{m}^{\mathrm{s}}}{1-\kappa} \right)}.
\end{align}
It is worth noting that $(\mathcal{R} _{\mathrm{u},\mathrm{c}}^{\mathrm{f}},\mathcal{R} _{\mathrm{u},\mathrm{s}}^{\mathrm{f}})$ can be discussed in the way we discuss $(\mathcal{R} _{\mathrm{u},\mathrm{c}}^{\mathrm{c}},\mathcal{R} _{\mathrm{u},\mathrm{s}}^{\mathrm{s}})$. Additionally, the rate region achieved by FDSAC is given by
\begin{align}
\mathcal{C} _{\mathrm{u},\mathrm{f}}\!=\!\left\{ \left( \mathcal{R} _{\mathrm{u},\mathrm{s}},\mathcal{R} _{\mathrm{u},\mathrm{c}} \right) \left|\!\! \begin{array}{c}
	\mathcal{R} _{\mathrm{u},\mathrm{s}}\!\in\! \left[ 0,\mathcal{R} _{\mathrm{u},\mathrm{s}}^{\mathrm{f}} \right] ,\mathcal{R} _{\mathrm{u},\mathrm{c}}\!\in\! \left[ 0,\mathcal{R} _{\mathrm{u},\mathrm{c}}^{\mathrm{f}} \right]\\
	,\kappa \in \left[ 0,1 \right]\\
\end{array} \right.\!\!\! \right\}  .    
\end{align}
Upon concluding all the analyses of uplink case, we consolidate the outcomes pertaining to diversity order and high-SNR slope within Table~\ref{table2}.

\begin{remark}\label{up_FDSAC_compare}
The results in Table~\ref{table2} suggest that the uplink ISAC and FDSAC achieve the same diversity order. However, ISAC achieves larger high-SNR slopes for both SR and CR compared to FDSAC, indicating that ISAC provides more DoFs in terms of both S\&C.
\end{remark}

\section{Numerical Results} \label{numerical}
\begin{figure*}[!t]
    \centering
    \subfigbottomskip=5pt
	\subfigcapskip=0pt
\setlength{\abovecaptionskip}{0pt}
    \subfigure[Sum ECRs.]
    {
        \includegraphics[width=0.295\textwidth]{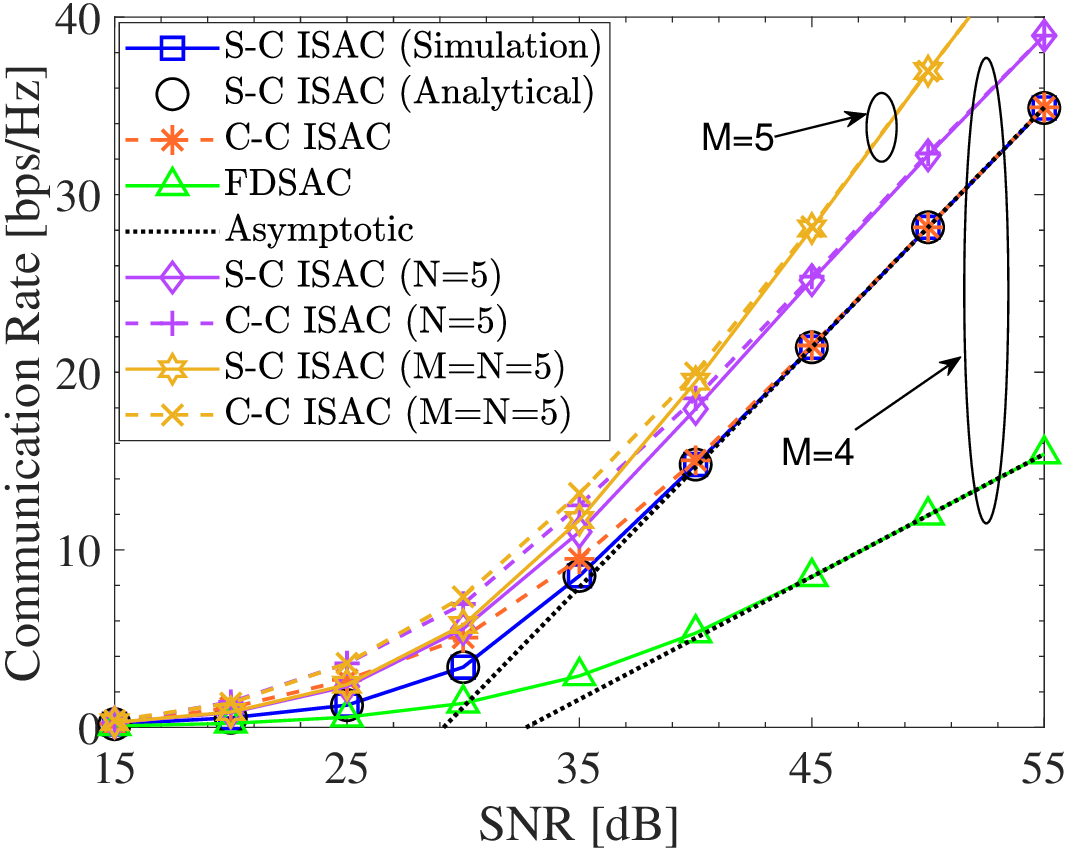}
	   \label{sum_ECR}	
    }
    \quad
   \subfigure[ECRs of UTs.]
    {
        \includegraphics[width=0.295\textwidth]{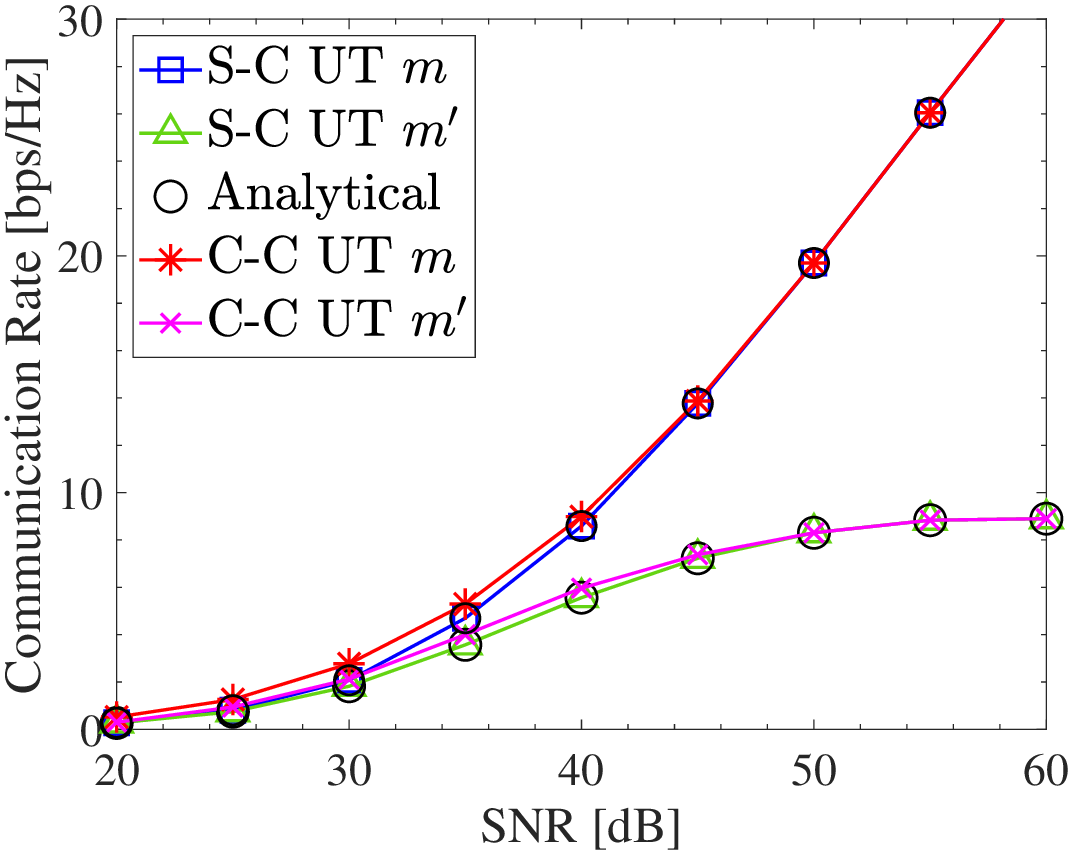}
	   \label{UT_ECR}	
    }
    \quad
   \subfigure[OPs with $\mathcal{R}_{\rm{d},m}^{\rm{t}}=\mathcal{R}_{\rm{d},m'}^{\rm{t}}=1$ bps/Hz.]
    {
        \includegraphics[width=0.305\textwidth]{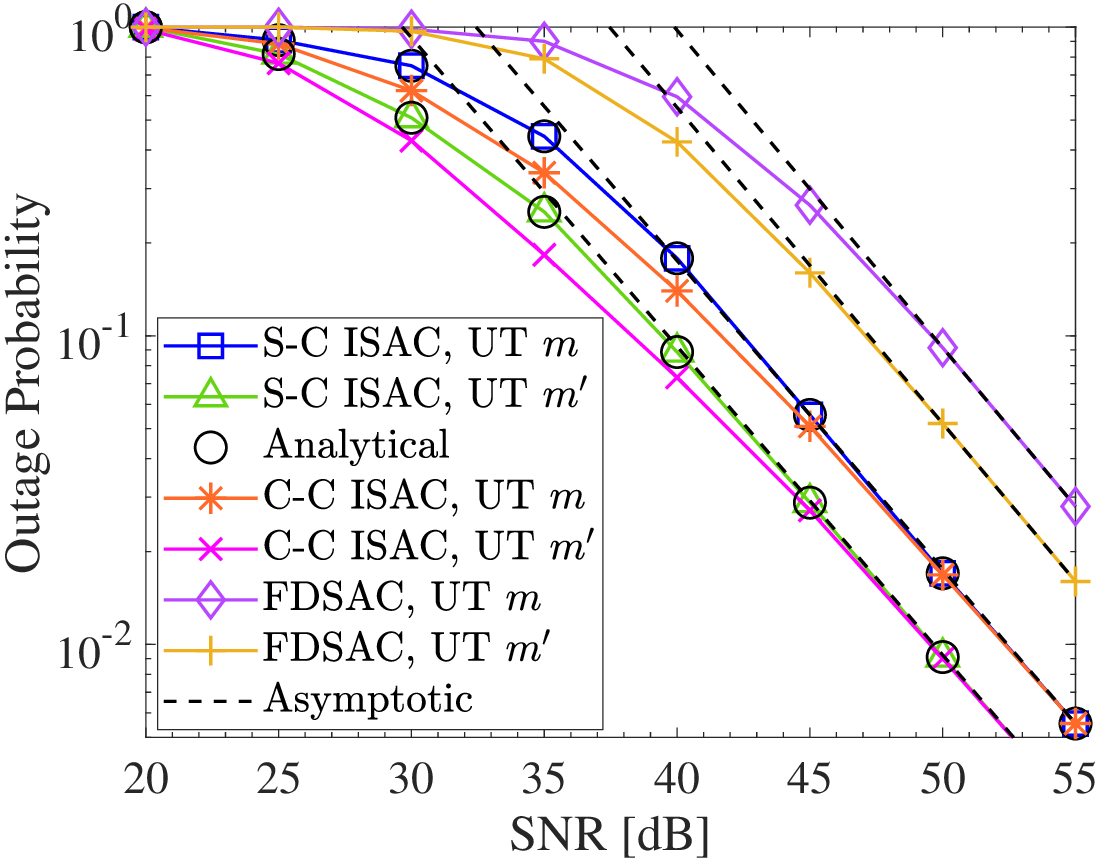}
	   \label{UT_OP}	
    }
\caption{Downlink communications performance.}
\vspace{-5pt}
    \label{do_com}
\end{figure*}

In this section, numerical results are provided to evaluate the S\&C performance of ISAC systems and verify the derived analytical results. Without otherwise specification, the simulation setups are outlined as follows. The antenna numbers of the BS and UTs are set as $M=4, N=4$, and the number of UTs is given by $2M=8$, i.e., $4$ pairs of UTs, each pair comprising a UT $m$ and UT $m'$ with $m\in \left\{ 1,2,3,4 \right\} $. In particular, UT $m$ and UT $m'$ are located on two circles centered at the base station, respectively. The NOMA power allocation coefficients are set as $\alpha_m=0.2$ and $\alpha_{m'}=0.8$, and the path loss are assumed to be $\eta_m^{-1}=1\times10^{-2}$ and $\eta_{m'}^{-1}=2.5\times10^{-3}$, for $m=1,2,3,4$. Besides, the length of the ISAC signal is $L=30$, the FDSAC parameters are $\kappa=\mu=0.5$, and the eigenvalues of $\mathbf{R}$ are $\{1,0.1,0.05,0.01\}$. 
\subsection{Downlink ISAC}
Fig.~\ref{sum_ECR} plots the downlink sum ECR versus the transmit SNR $p$. It can be seen that C-C ISAC achieves the best performance while FDSAC achieves the lowest sum ECR. The derived analytical results of sum ECR under the S-C design match the simulation results well, while the asymptotic results precisely track the simulation results in the high-SNR region. Furthermore, the sum ECRs under various antenna numbers are also presented. By increasing the number of the UT's antenna to $5$, the communication rate is obviously improved in both the S-C design and the C-C design, while the high-SNR slopes remain the same. When the number of the BS antennas increases, e.g., $M=N=5$ with an additional eigenvalue of $0.3$ for $\mathbf{R}$, more UTs can be served. In this case, in addition to the growth in the sum ECR, the high SNR slopes are also enhanced. These observations align with the derived high-SNR slope, which equals $M$. 

In Fig.~\ref{UT_ECR}, the ECR of UT $m$ and $m'$ are respectively illustrated. We can observe that the ECR of UT $m$ increases linearly when SNR goes high, while UT $m'$ converges to a upper bound. Fig.~\ref{UT_OP} displays the downlink outage performance, confirming the validity of the analytical and asymptotic results. From Fig.~\ref{do_com}, we find that the S-C ISAC and the C-C ISAC have the same high-SNR performance that is change linearly with the SNR, which is consistent with the statement in \textbf{Remark~\ref{CR_compare}}.

\begin{figure} [t!]
\centering
\includegraphics[width=0.32\textwidth]{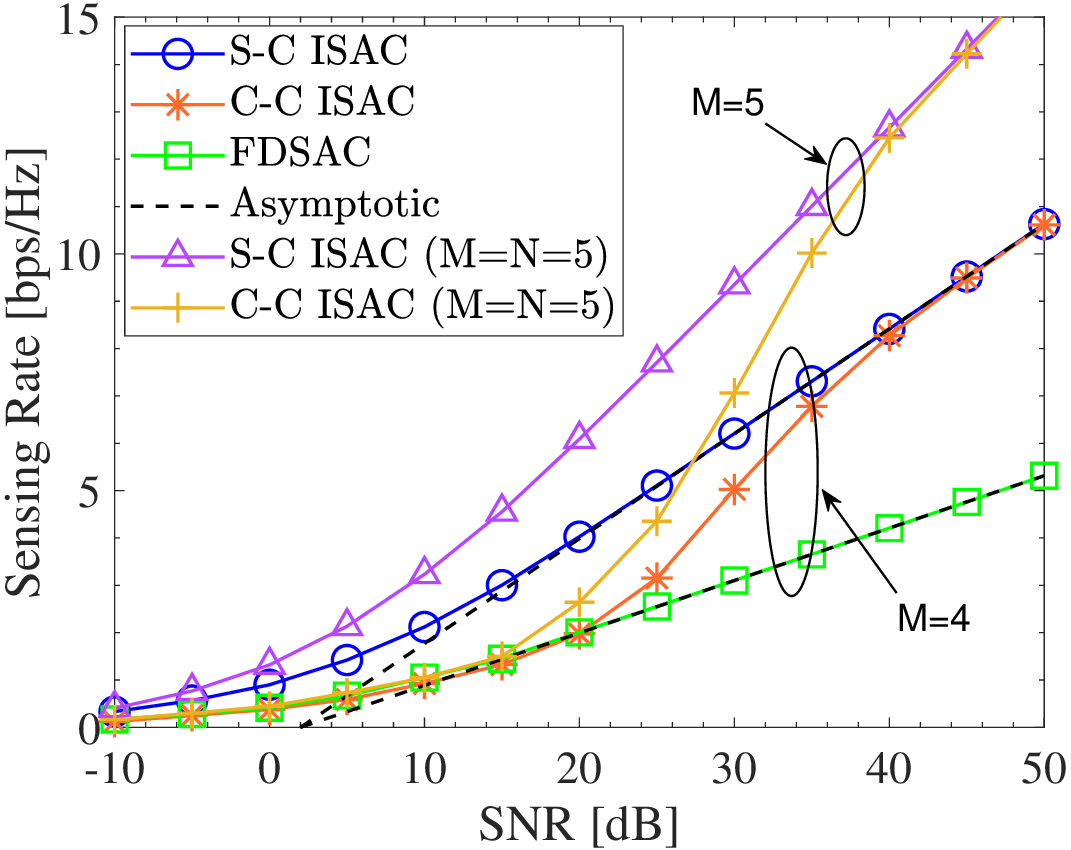}
 \caption{Downlink sensing rate.}
 \vspace{-5pt}
 \label{SR}
\end{figure}

Fig.~\ref{SR} shows the downlink sensing performance. We can observe that S-C ISAC exhibits the most superior SR performance, while the performance achieved by FDSAC is still the worst. We evaluate the sensing performance with varying antenna numbers of the BS. When $M$ is increased to $5$, both sensing rates and high-SNR slopes improve. It is also noticeable that the performance gap between the S-C and the C-C design becomes more significant as the number of BS antennas increases. By jointly examining Fig.~\ref{do_com} and Fig.~\ref{SR} together, it can be observed that in high SNR regime, both downlink communication and sensing performance of the S-C ISAC and the C-C ISAC exhibit similarities and demonstrate linear improvement with increasing SNR, which corroborates the conclusion drawn in \textbf{Remark~\ref{Sc_CC_compare}}. Moreover, it is evident that ISAC achieves larger high-SNR slopes than FDSAC in terms of both downlink CR and SR, while they achieve the same diversity order, verifying the correctness of \textbf{Remark~\ref{FDSAC_compare}}.

\begin{figure} [t!]
\centering
\includegraphics[width=0.32\textwidth]{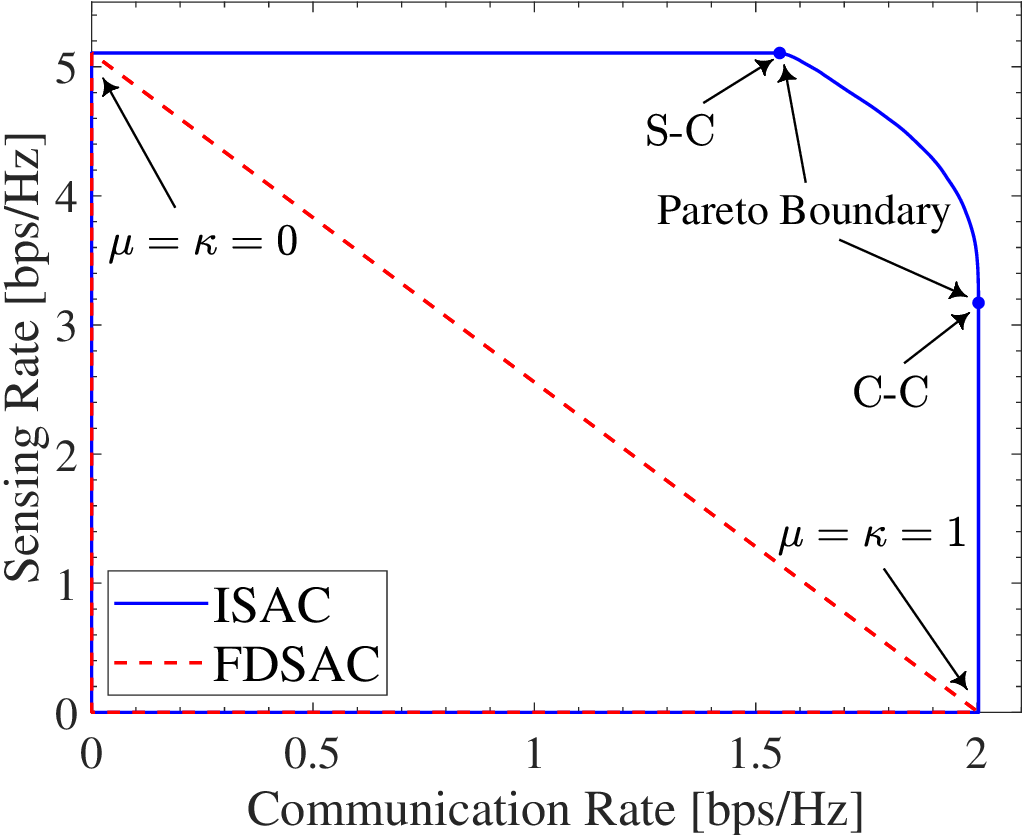}
 \caption{Downlink rate region with $p=25$ dB.}
 \vspace{-5pt}
 \label{region}
\end{figure}

\begin{figure}[!t]
    \centering
    \subfigbottomskip=5pt
	\subfigcapskip=0pt
\setlength{\abovecaptionskip}{0pt}
    \subfigure[Sum ECRs.]
    {
        \includegraphics[width=0.315\textwidth]{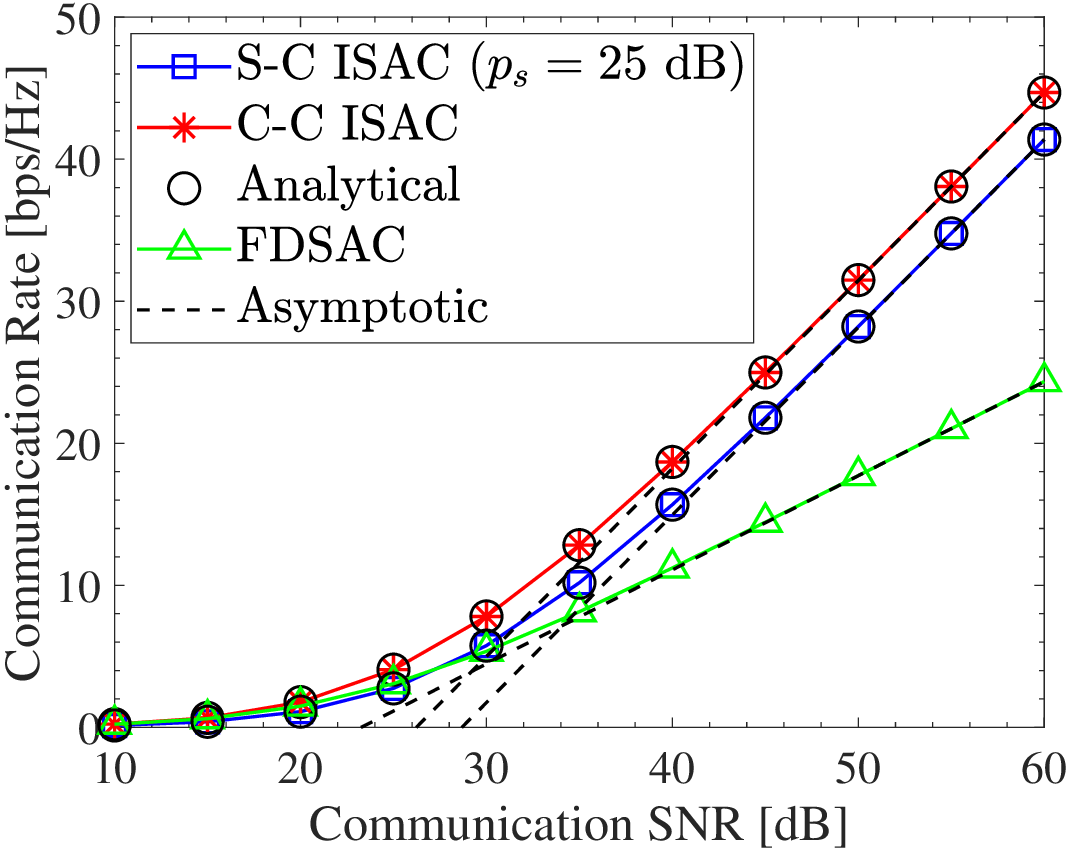}
	   \label{up_CR}	
    }
   \subfigure[OPs of a UT pair.]
    {
        \includegraphics[width=0.32\textwidth]{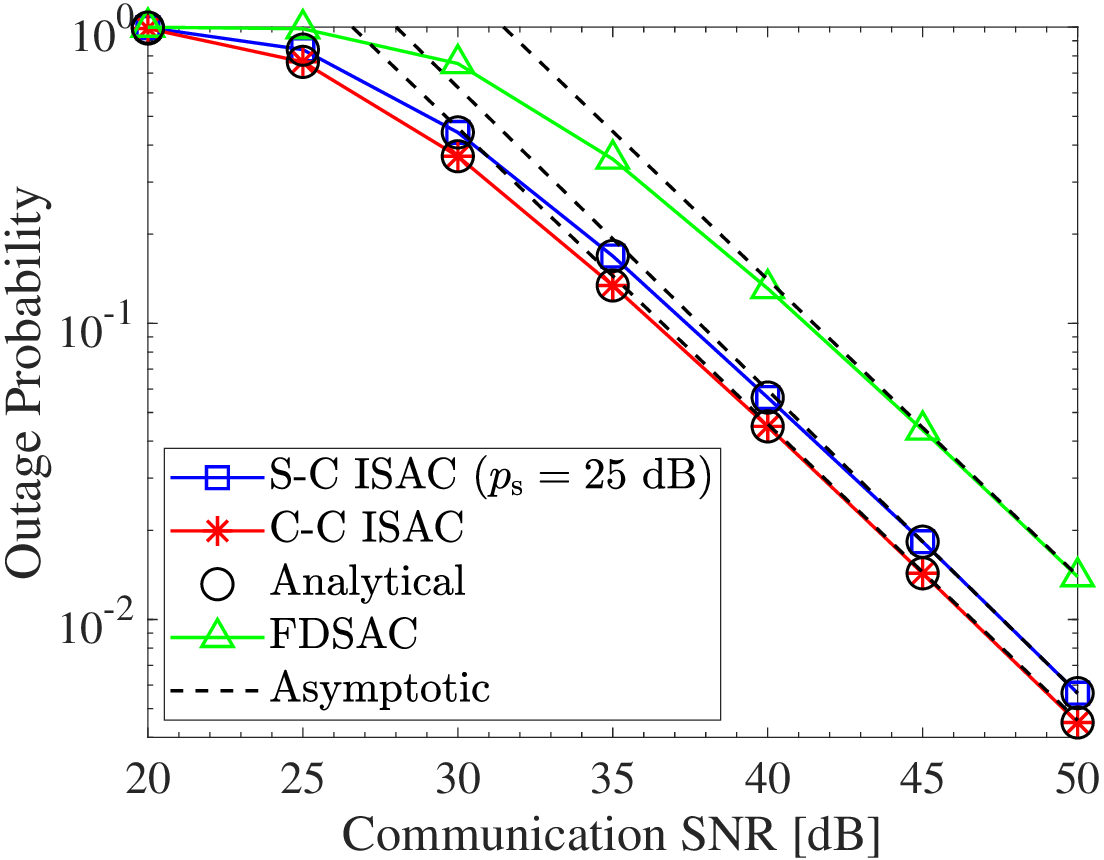}
	   \label{up_OP}	
    }
\caption{Uplink communications performance.}
\vspace{-5pt}
    \label{up_communications}
\end{figure}

In Fig.~\ref{region}, the downlink SR-CR regions attained by the two systems are displayed: ISAC system (presented in \eqref{Rate_Regio_ISAC}) and the baseline FDSAC system (presented in \eqref{Rate_Regio_FDSAC}). The two marked points on the graph represent the S-C and C-C designs, respectively. The curve section connecting these two points represents the Pareto boundary of downlink ISAC’s rate region, which was derived by solving the problem \eqref{Problem_CR_SR_Tradeoff} for values of $\rho$ ranging from $1$ to $0$. The rate region of FDSAC is plotted by changing both the bandwidth allocation factor $\kappa$ and power allocation factor $\mu$ from $0$ to $1$. Specifically, we iterate over $\mu$ from $0$ to $1$ with a step size $0.01$. During this process, for each given $\mu$, $\kappa$ iterates from $0$ to $1$ also with a step size $0.01$. As expected, larger values of $\kappa$ and $\mu$ yield a higher CR. It is noteworthy that the rate region achieved by downlink FDSAC is entirely contained within that of downlink ISAC, justifying the correctness of \textbf{Theorem~\ref{rate_region}}.

\begin{figure} [t!]
\centering
\includegraphics[width=0.315\textwidth]{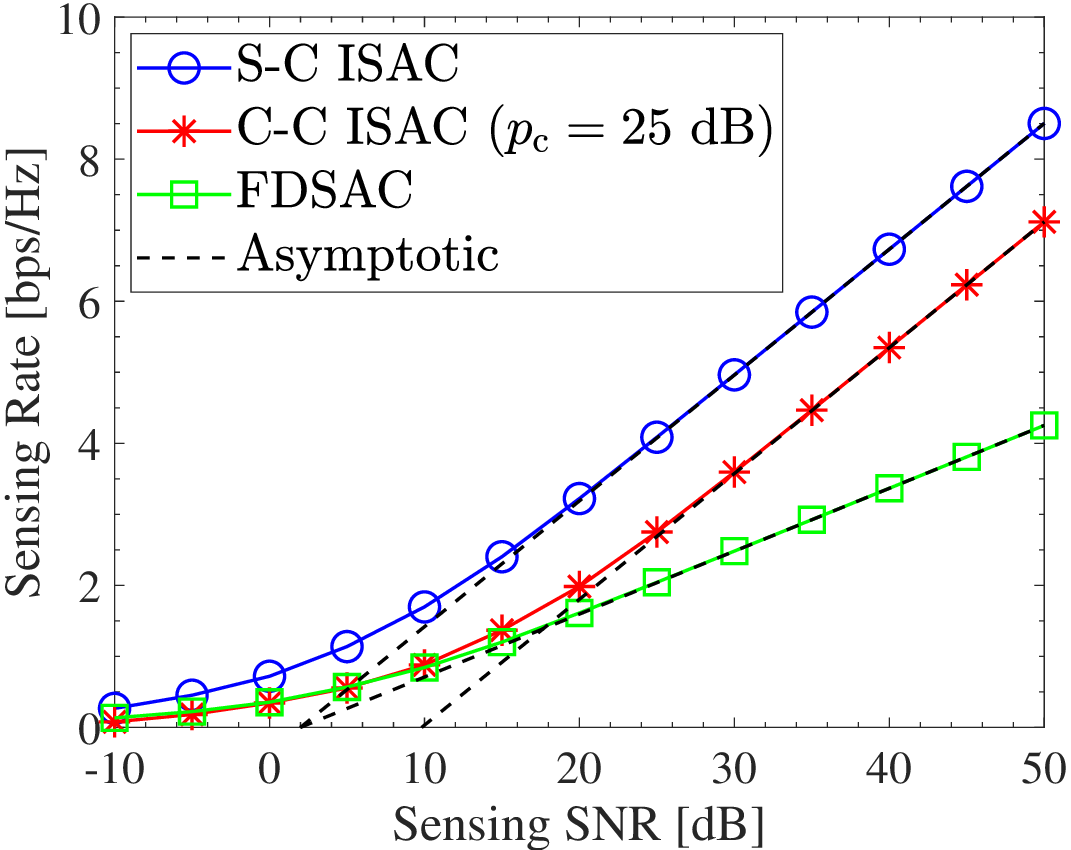}
 \caption{Uplink sensing rate.}
 \vspace{-2pt}
 \label{up_SR}
\end{figure}

\begin{figure} [t!]
\centering
\includegraphics[width=0.32\textwidth]{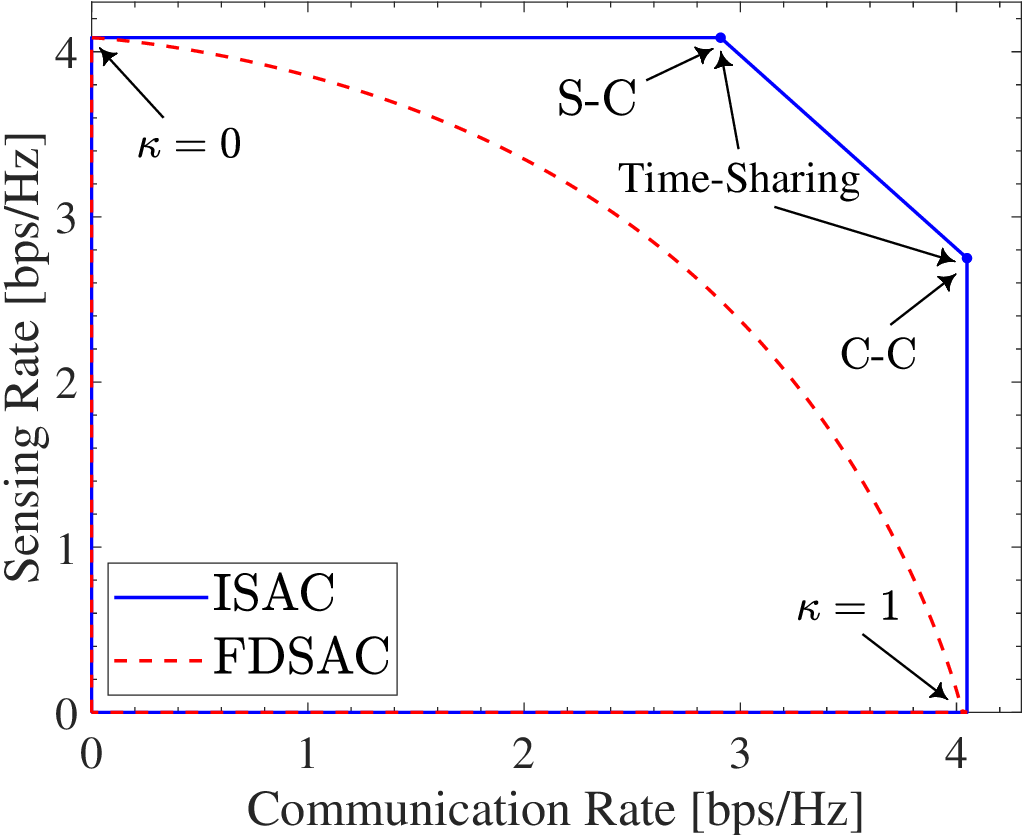}
 \caption{Uplink rate region with $p_{\rm{c}}=p_{\rm{u}}=25$ dB.}
 \vspace{-5pt}
 \label{up_region}
\end{figure}
\subsection{Uplink ISAC}
Then moving on to the uplink results, we examine Fig.~\ref{up_CR} and Fig.~\ref{up_OP}, which depict the uplink sum rate and outage performance with $\mathcal{R}_{\rm{u}}^{\rm{t}}=2$ bps/Hz concerning the communication SNR $p_{\rm{c}}$, respectively. Notably, the C-C ISAC exhibits the highest communication performance, while the FDSAC demonstrates the lowest performance. The analytical results align well with the simulation results, and the asymptotic results accurately capture the behavior in the high-SNR regime. It is noteworthy that C-C ISAC and S-C ISAC achieve identical high-SNR slopes and diversity orders, which aligns with the our former discussions in Section~\ref{Section: Discussion on SIC Ordering}. Moreover, the high-SNR slope achieved by the uplink ISAC is larger than that achieved by FDSAC, while their diversity orders are same, as highlighted in \textbf{Remark~\ref{up_FDSAC_compare}}.

Fig.~\ref{up_SR} shows the uplink SR versus sensing SNR $p_{\rm{s}}$. As anticipated, S-C ISAC achieves the highest SR, while FDSAC achieves the lowest. We observe that the asymptotic results accurately track the provided simulation results in the high-SNR regime. Besides, C-C ISAC and S-C ISAC achieve the same high-SNR slope that is larger than that achieved by FDSAC. Furthermore, it is noteworthy that when achieving the same SR in the high-SNR region, S-C ISAC outperforms C-C ISAC by a constant SNR gap. This observation highlights the superiority of S-C design over C-C design in terms of spectral efficiency in the high-SNR regime. In Fig.~\ref{up_region}, the achieved SR-CR regions by the uplink FDSAC and ISAC are presented. In the case of ISAC, the two points represent the rates achieved by the S-C and C-C schemes, respectively, while the line segment connecting these points represents the rates achievable through time-sharing strategy between the two schemes. A crucial observation from the plot is that the rate region of the uplink FDSAC is entirely contained within the rate region of the uplink ISAC, which clearly demonstrates the superiority of ISAC over FDSAC.

\section{Conclusion}\label{conclusion}
This paper has investigated the S\&C performance of a MIMO-based NOMA-ISAC framework for both downlink and uplink cases. The concept of signal alignment was utilized to efficiently serve a larger number of UTs. The downlink ISAC was analyzed under three typical scenarios: S-C design, C-C design and Pareto optimal design, while two different scenario based on the SIC order were considered in uplink case. For each scenario, SRs, CRs and OPs are derived. To gain a better insight into the S\&C performance of ISAC, we also investigated the asymptotic performance in high-SNR regime, including high-SNR slopes and diversity orders. In addition, the downlink and uplink SR-CR rate regions achieved by ISAC and the conventional FDSAC were characterized. The results have demonstrated that ISAC can achieve a more extensive rate region compared to FDSAC in both downlink and uplink cases, providing more DoFs and highlighting its superiority in terms of S\&C performance.

\begin{appendices}
\section{The Relationship Between Sensing MI and Estimation Error}\label{Appendix:0}
\renewcommand{\theequation}{A.\arabic{equation}}
\setcounter{equation}{0}
We consider the sensing model given in \eqref{reflected_echo_signal_matrix}. Under our considered system, the response matrix can be equivalently written as $\mathbf{G}^{\mathsf{H}}=\mathbf{QR}^{\frac{1}{2}}$, where $\mathbf{Q}\in{\mathbbmss{C}}^{M\times M}$ contains $M^2$ independent and identically distributed elements with zero mean and unit variance. For a given waveform matrix $\mathbf{X}$, the resultant MSE in estimating the target response matrix $\mathbf{G}$ can be expressed as follows \cite{MSE}: 
\setlength\abovedisplayskip{4pt}
\setlength\belowdisplayskip{4pt}
\begin{align}
\mathsf{MSE}=M\mathsf{tr}\left( \left( \mathbf{XX}^{\mathsf{H}}+\mathbf{R}^{-1} \right) ^{-1} \right) .
\end{align}
Consequently, the optimal waveform in minimizing the MSE is given by
\begin{align}
\mathbf{X}_{1}^{\star}=\argmax\nolimits_{\mathsf{tr}\left( \mathbf{XX}^{\mathsf{H}} \right) \le Lp}\mathsf{tr}\left( \left( \mathbf{XX}^{\mathsf{H}}+\mathbf{R}^{-1} \right) ^{-1} \right).    
\end{align}
Based on \cite{MSE}, the optimal solution to the above problem satisfies
\begin{align}
\mathbf{X}_{1}^{\star}{\mathbf{X}_{1}^{\star}}^{\mathsf{H}}=\mathbf{U}^{\mathsf{H}}\mathsf{diag}\left\{ s_{1}^{\star},\ldots,s_{M}^{\star} \right\} \mathbf{U}.
\end{align}
Moreover, for a given waveform matrix $\mathbf{X}$, the MI between $\mathbf{G}$ and $\mathbf{Y}_\mathrm{s}$ is given by \cite{ouyang2022integrated}
\begin{align}
\mathcal{I} =M\log _2\det \left( \mathbf{I}_M+\mathbf{X}^{\mathsf{H}}\mathbf{RX} \right) .
\end{align}
By definition, the sensing rate can be written as
\begin{align}
\mathcal{R} _{\mathrm{s}}=\frac{M}{L}\log _2\det \left( \mathbf{I}_M+\mathbf{X}^{\mathsf{H}}\mathbf{RX} \right). 
\end{align}
Consequently, the optimal waveform in terms of maximizing the sensing rate is given by
\begin{align}
\mathbf{X}_{2}^{\star}=\argmax\nolimits _{\mathsf{tr}\left( \mathbf{XX}^{\mathsf{H}} \right) \le Lp}\log _2\det \left( \mathbf{I}_M+\mathbf{X}^{\mathsf{H}}\mathbf{RX} \right) .
\end{align}
Based on the derivations in Appendix \ref{Proof_Sensing_Rate_S_C_Theorem}, the optimal solution to the above problem satisfies
\begin{align}
\mathbf{X}_{2}^{\star}{\mathbf{X}_{2}^{\star}}^{\mathsf{H}}=\mathbf{U}^{\mathsf{H}}\mathsf{diag}\left\{ s_{1}^{\star},\ldots,s_{M}^{\star} \right\} \mathbf{U}.
\end{align}
Comparing $\mathbf{X}_{1}^{\star}$ with $\mathbf{X}_{2}^{\star}$, we can conclude that the waveform that can maximize the sensing rate can also minimize the MSE in estimating the target response matrix $\mathbf{G}$. Or in other words, maximizing the sensing rate is equivalent to minimizing the MSE in estimating the target response matrix $\mathbf{G}$. It is worth noting that the detailed discussions about the relationship between the sensing rate and the mean-square error have been widely reported; see \cite{ouyang2022integrated,MSE,Tang2019_TSP} and the references therein.

\section{Proof of Lemma \ref{SR_Basic_Lemma}}\label{Appendix:A}
\renewcommand{\theequation}{B.\arabic{equation}}
\setcounter{equation}{0}
Based on \cite{Tang2019_TSP}, the sensing MI is calculated as follows: 
\begin{align}
 I\left({\mathbf{Y}}_{\rm{s}};{\mathbf{G}}|{\mathbf{X}}\right)=M\log_2\det\left({\mathbf{I}}_{M}
+{\mathbf{X}}^{\mathsf{H}}{\mathbf{R}}{\mathbf{X}}\right).   
\end{align} 
Applying the Sylvester's identity and the fact of $\mathbf{X}^{\mathsf{H}}\mathbf{X}=L\mathbf{P}^\mathsf{H}\mathbf{P}$, we can rewrite the MI as 
\begin{align}
I\left( \mathbf{Y}_{\mathrm{s}};\mathbf{G}|\mathbf{X} \right) =M\log _2\det\left(\mathbf{I}_M+L\mathbf{P}^\mathsf{H}\mathbf{R}\mathbf{P}\right).
\end{align}
Then based on \eqref{eq_SR}, we obtain the downlink SR in Lemma~\ref{SR_Basic_Lemma}.

\section{Proof of Theorem \ref{Sensing_Rate_S_C_Theorem}}\label{Proof_Sensing_Rate_S_C_Theorem}
\renewcommand{\theequation}{C.\arabic{equation}}
\setcounter{equation}{0}
It is noteworthy that maximizing the SR ${\mathcal{R}}_{\rm{d},\rm{s}}=L^{-1}M\log_2\det\left({\mathbf{I}}_M+L{\mathbf{P}}^{\mathsf{H}}{\mathbf{R}}{\mathbf{P}}\right)$ is equivalent to maximizing the MI of a virtual MIMO channel:
$\dot{\mathbf{y}}={\mathbf{R}}^{\frac{1}{2}}\dot{\mathbf{x}}+\dot{\mathbf{n}}$, where ${\mathbb{E}}\{\dot{\mathbf{x}}{\dot{\mathbf{x}}}^{\mathsf{H}}\}={\mathbf{P}}{\mathbf{P}}^{\mathsf{H}}$ and $\dot{\mathbf{n}}\sim{\mathcal{CN}}\left({\mathbf{0}},L^{-1}{\mathbf{I}}_M\right)$. Consequently, when ${\mathcal{R}}_{\rm{d},\rm{s}}$ is maximized, the eigenvectors of ${\mathbf{P}}{\mathbf{P}}^{\mathsf{H}}$ should align with the eigenvectors of ${\mathbf{R}}^{\frac{1}{2}}$, while the eigenvalues are determined using the water-filling procedure \cite{mimo}, which yields 
\begin{align}
\mathcal{R}_{\rm{d},\rm{s}}^{\rm{s}}=\frac{M}{L}\sum\nolimits_{m=1}^{M}\log_2\left(1+L\lambda_ms_m^{\star}\right),   
\end{align}
where $s_{m}^{\star}=\max \{0,\frac{1}{\nu}-\frac{1}{L\lambda _m}\}$ and $\sum_{m=1}^{M}s_{m}^{\star}=p$. 

When $p\rightarrow\infty$, we have $s_{m}^{\star}=\frac{1}{\nu}-\frac{1}{L\lambda _m}$ for all $m$, and thus the equations $\sum_{m=1}^M{s_{m}^{\star}}=\frac{M}{\nu}-\sum_{m=1}^M{\frac{1}{L\lambda _m}}=p$ hold, which yields
\begin{align}\label{b4}
s_{m}^{\star}=\frac{p}{M}-\frac{1}{L\lambda _m}+\frac{1}{LM}\sum\nolimits_{m=1}^M{\frac{1}{\lambda _m}}.
\end{align}
By substituting \eqref{b4} into \eqref{Sensing_Rate_S_C_exact} and applying the fact of $\lim_{x\rightarrow\infty}\log_2(a+x)\approx\log_2{x}$, the result in \eqref{Sensing_Rate_S_C_Asymptotic} is obtained.

\section{Proof of Lemma \ref{SC_ECR}}\label{Proof_SC_ECR_Lemma}
\renewcommand{\theequation}{D.\arabic{equation}}
\setcounter{equation}{0}
Since we have $\mathbf{H}_{\mathrm{d},m}^{\mathsf{H}}\mathbf{v}_{\mathrm{d},m}=\mathbf{u}_m$, we can obtain $\mathbf{v}_{\mathrm{d},m}^{\mathsf{H}}\mathbf{H}_{\mathrm{d},m}\mathbf{H}_{\mathrm{d},m}^{\mathsf{H}}\mathbf{v}_{\mathrm{d},m}=\mathbf{u}_{m}^{\mathsf{H}}\mathbf{u}_m=1$. Performing ED on the complex Wishart matrix $\mathbf{H}_{\mathrm{d},m}\mathbf{H}_{\mathrm{d},m}^{\mathsf{H}}$, we get $\mathbf{v}_{\mathrm{d},m}^{\mathsf{H}}\mathbf{Q\Lambda Q}^{-1}\mathbf{v}_{\mathrm{d},m}=1\leqslant \mathbf{v}_{\mathrm{d},m}^{\mathsf{H}}\mathbf{Q}\lambda _{m}^{\max}\mathbf{IQ}^{-1}\mathbf{v}_{\mathrm{d},m}=\lambda _{m}^{\max}\left\| \mathbf{v}_{\mathrm{d},m} \right\|^2$, where $\lambda _{m}^{\max}$ denotes the largest eigenvalue of $\mathbf{H}_{\mathrm{d},m}\mathbf{H}_{\mathrm{d},m}^{\mathsf{H}}$. Therefore, a lower bound of $\left\| \mathbf{v}_{\mathrm{d},m} \right\|^2$ is obtained as $\left\| \mathbf{v}_{\mathrm{d},m} \right\|^2\geqslant \frac{1}{\lambda _{m}^{\max}}$. Similarly, we can get $\left\| \mathbf{v}_{\mathrm{d},m'} \right\|^2\geqslant \frac{1}{\lambda _{m'}^{\max}}$, where $\lambda _{m'}^{\max}$ is the largest eigenvalue of $\mathbf{H}_{\mathrm{d},m'}\mathbf{H}_{\mathrm{d},m'}^{\mathsf{H}}$. By replacing $\left\| \mathbf{v}_{\mathrm{d},m} \right\|^2$ and $\left\| \mathbf{v}_{\mathrm{d},m'} \right\|^2$ with their lower bounds $\frac{1}{\lambda _{m}^{\max}}$ and $\frac{1}{\lambda _{m'}^{\max}}$, respectively, we can obtain the upper bounds of $\mathcal{R} _{\mathrm{d},m}^{\mathrm{s}}$ and $ \mathcal{R} _{\mathrm{d},m'}^{\mathrm{s}}$ as follows:
\begin{align}
\tilde{\mathcal{R}}_{\mathrm{d},m}^{\mathrm{s}}&=\mathbb{E} \left\{ \log _2\left( 1+\alpha _m\eta _{m}^{-1}s_{m}^{\star}\lambda _{m}^{\max} \right) \right\} \notag\\
&=\int_0^{\infty}{\log _2\left( 1+\alpha _m\eta _{m}^{-1}s_{m}^{\star}x \right) f_{\gamma}\left( x \right) dx},\\
\tilde{\mathcal{R}}_{\mathrm{d},m'}^{\mathrm{s}}&=\mathbb{E} \left\{ \log _2\left( 1+\frac{\alpha _{m^{\prime}}\eta _{m^{\prime}}^{-1}s_{m}^{\star}\rho _{m^{\prime}}^{2}\lambda _{m^{\prime}}^{\max}}{\alpha _m\eta _{m}^{-1}s_{m}^{\star}\rho _{m}^{2}\lambda _{m^{\prime}}^{\max}+1} \right) \right\} \notag\\
&=\int_0^{\infty}{\log _2\left( 1+\frac{\alpha _{m^{\prime}}\eta _{m^{\prime}}^{-1}s_{m}^{\star}x}{\alpha _m\eta _{m^{\prime}}^{-1}s_{m}^{\star}x+1} \right) f_{\gamma}\left( x \right) dx}.
\end{align}

When $N=M$, since $\left\| \mathbf{v}_{\mathrm{d},m} \right\|^2$ can be analytically expressed as $\mathbf{v}_{\mathrm{d},m}= \mathbf{H}_{\mathrm{d},m}^{-\mathsf{H}} \mathbf{u}_m$, we have $\left\| \mathbf{v}_{\mathrm{d},m} \right\|^2=\mathbf{u}_{m}^{\mathsf{H}}\mathbf{H}_{m}^{-1}\mathbf{H}_{m}^{-\mathsf{H}}\mathbf{u}_m=\left( \mathbf{U}^\mathsf{H}\mathbf{H}_{m}^{-1}\mathbf{H}_{m}^{-\mathsf{H}}\mathbf{U} \right) _{m,m}$. As $\mathbf{U}$ is a unitary matrix, the statistical properties of $\left( \mathbf{U}^\mathsf{H}\mathbf{H}_{m}^{-1}\mathbf{H}_{m}^{-\mathsf{H}}\mathbf{U} \right) _{m,m}$ are same as $\left( \mathbf{H}_{m}^{-1}\mathbf{H}_{m}^{-\mathsf{H}} \right) _{m,m}$. Therefore, $\frac{1}{\left\| \mathbf{v}_{\mathrm{d},m} \right\|^2}$ has the same statistical properties as $\frac{1}{\left( \mathbf{H}_{m}^{-1}\mathbf{H}_{m}^{-\mathsf{H}} \right) _{m,m}}$, which is exponentially distributed \cite{Globe_2003}. Similarly, $\frac{1}{\left\| \mathbf{v}_{\mathrm{d},m'} \right\|^2}$ is also exponentially distributed. As a result, the closed-form expressions of $\mathcal{R} _{\mathrm{d},m}^{\mathrm{s}}$ and $ \mathcal{R} _{\mathrm{d},m'}^{\mathrm{s}}$ can be derived as follows:
\begin{align}
\mathcal{R} _{\mathrm{d},m}^{\mathrm{s}}&=\int_0^{\infty}{\log _2\left( 1+\alpha _m\eta _{m}^{-1}s_{m}^{\star}x \right) e^{-x}dx},\\
\mathcal{R} _{\mathrm{d},m'}^{\mathrm{s}}&=\int_0^{\infty}{\log _2\left( 1+\left( \alpha _m\eta _{m^{\prime}}^{-1}s_{m}^{\star}+\alpha _{m^{\prime}}\eta _{m^{\prime}}^{-1}s_{m}^{\star} \right) x \right) e^{-x}dx}\notag\\
&~~~-\int_0^{\infty}{\log _2\left( 1+\alpha _m\eta _{m^{\prime}}^{-1}s_{m}^{\star}x \right)\! e^{-x}dx}.
\end{align}
Then, with the aid of \cite[Eq. (4.337.2)]{integral}, the results in \eqref{SC_CR_M} and \eqref{SC_CR_M'} can be derived.

\section{Proof of Theorem \ref{SC_OP}}\label{Proof_SC_OP_theorem}
\renewcommand{\theequation}{E.\arabic{equation}}
\setcounter{equation}{0}
The lower bounds for $\mathcal{P} _{\mathrm{d},m}^{\mathrm{s}}$ and $\mathcal{P} _{\mathrm{d},m'}^{\mathrm{s}}$ are derived by replacing $\left\| \mathbf{v}_{\mathrm{d},m} \right\|^2$ and $\left\| \mathbf{v}_{\mathrm{d},m'} \right\|^2$ with $\frac{1}{\lambda _{m}^{\max}}$ and $\frac{1}{\lambda _{m'}^{\max}}$, respectively:
\begin{align}
\tilde{\mathcal{P}}_{\mathrm{d},m}^{\mathrm{s}}&=1-\mathrm{Pr}\left( \lambda _{m}^{\max}>\frac{\varrho _m}{s_{m}^{\star}} \right) \mathrm{Pr}\left( \lambda _{m^{\prime}}^{\max}>\frac{\varrho _{m'}}{s_{m}^{\star}} \right)\notag\\
&=1-\left( 1-F_{\lambda}\left( \frac{\varrho _m}{s_{m}^{\star}} \right) \right) \left( 1-F_{\lambda}\left( \frac{\varrho _{m'}}{s_{m}^{\star}} \right) \right) ,\\
\tilde{\mathcal{P}}_{\mathrm{d},m^\prime}^{\mathrm{s}}&=\mathrm{Pr}\left( \lambda _{m^{\prime}}^{\max}<\frac{\varrho _{m'}}{s_{m}^{\star}} \right) =F_{\lambda}\left( \frac{\varrho _{m'}}{s_{m}^{\star}} \right) .
\end{align}
When $N=M$, as stated before, $\frac{1}{\left\| \mathbf{v}_{\mathrm{d},m} \right\|^2}$ and $\frac{1}{\left\| \mathbf{v}_{\mathrm{d},m'} \right\|^2}$ follow the standard exponential distributions. Hence, we can derive the closed-form expressions of $\mathcal{P} _{\mathrm{d},m}^{\mathrm{s}}$ and $\mathcal{P} _{\mathrm{d},m'}^{\mathrm{s}}$ as follows:
\begin{align}
\mathcal{P} _{\mathrm{d},m}^{\mathrm{s}}&=1-\left( 1-F_v\left( \frac{\varrho _{m}}{s_{m}^{\star}} \right) \right) \left( 1-F_v\left( \frac{\varrho _{m'}}{s_{m}^{\star}} \right) \right) ,\\
\mathcal{P} _{\mathrm{d},m^\prime}^{\mathrm{s}}&=F_v\left( \frac{\varrho _{m'}}{s_{m}^{\star}} \right) ,
\end{align}
where $F_v\left( x \right) =1-\exp \left( -x \right) $ denotes the CDF of $\frac{1}{\left\| \mathbf{v}_{\mathrm{d},m} \right\|^2}$ and $\frac{1}{\left\| \mathbf{v}_{\mathrm{d},m'} \right\|^2}$, enabling the derivation of results upon substitution.

\section{Proof of Theorem \ref{rate_region}}\label{Proof_rate_region}
\renewcommand{\theequation}{F.\arabic{equation}}
\setcounter{equation}{0}
Firstly, we define two auxiliary regions in the following manner:
\begin{align}
\mathcal{C} _1&=\left\{ \left( \mathcal{R} _{\mathrm{s}},\mathcal{R} _{\mathrm{c}} \right) |\mathcal{R} _{\mathrm{s}}\!\in \!\left[ 0,\mathcal{R} _{\mathrm{s},1}^{\epsilon} \right] ,\mathcal{R} _{\mathrm{c}}\!\in \!\left[ 0,\mathcal{R} _{\mathrm{c},1}^{\epsilon} \right] ,\epsilon \!\in \!\left[ 0,\!1 \right] \right\} ,\\
\mathcal{C} _2&=\left\{ \left( \mathcal{R} _{\mathrm{s}},\mathcal{R} _{\mathrm{c}} \right) |\mathcal{R} _{\mathrm{s}}\!\in \!\left[ 0,\mathcal{R} _{\mathrm{s},2}^{\epsilon} \right] ,\mathcal{R} _{\mathrm{c}}\!\in \!\left[ 0,\mathcal{R} _{\mathrm{c},2}^{\epsilon} \right] ,\epsilon \!\in \!\left[ 0,\!1 \right] \right\} ,
\end{align}
where $\mathcal{R} _{\mathrm{s},1}^{\epsilon}$ and $\mathcal{R} _{\mathrm{c},1}^{\epsilon}$ are defined as follows:
\begin{align}
\mathcal{R} _{\mathrm{s},1}^{\epsilon}\!&=\!\frac{M}{L}\!\max _{\sum_{m=1}^M{k_m}\le (1-\epsilon )p}\sum\nolimits_{m=1}^M\!{\log _2\left( 1\!+\!L\lambda _mk_m \right)},\\
\mathcal{R} _{\mathrm{c},1}^{\epsilon}\!&=\!\mathbb{E} \left\{ \max _{\sum_{m=1}^M{j_m}\le \epsilon p}\sum\nolimits_{m=1}^M{\log _2\left( 1\!+\!\frac{\alpha _m\eta _{m}^{-1}j_m}{\left\| \mathbf{v}_{\mathrm{d},m} \right\|^2} \right)} \right. \notag\\
&~~~~~~~~~~~\left.+\!\log _2\!\left( 1\!+\!\frac{\alpha _{m^{\prime}}\eta _{m^{\prime}}^{-1}j_m}{\alpha _m\eta _{m^{\prime}}^{-1}j_m\!+\!\left\| \mathbf{v}_{\mathrm{d},m'} \right\|^2} \right) \!\right\}.
\end{align}
For a given $\epsilon$, we define $k_{m}^{\epsilon}$ and $j_{m}^{\epsilon}$ as the optimal solutions for $m=1,\ldots,M$. Then, $\mathcal{R} _{\mathrm{s},2}^{\epsilon}$ and $\mathcal{R} _{\mathrm{c},2}^{\epsilon}$ represent the average SR and sum ECR achieved by the precoding matrix $\mathbf{P}_{\epsilon}=\mathbf{U}\bm{\Delta} _{\epsilon}^{\frac{1}{2}}$, where $\bm{\Delta} _{\epsilon}=\mathsf{diag}\left\{ k_{1}^{\epsilon}+j_{1}^{\epsilon},\ldots,k_{M}^{\epsilon}+j_{M}^{\epsilon} \right\} $. It is evident that $\mathcal{C} _{2}\subseteq \mathcal{C} _{\rm{d},\mathrm{i}}$ and $\mathcal{C} _{\rm{d},\mathrm{f}}\subseteq \mathcal{C} _{\mathrm{1}}$.

Given $\epsilon_1\in\left[0,1\right]$. When $\mathcal{R} _{\mathrm{c},1}^{\epsilon_1}\in \left[ 0, \mathcal{R} _{\mathrm{d},\mathrm{c}}^{\mathrm{s}}\right]$, we have $\mathcal{R} _{\mathrm{c},1}^{\epsilon_1}\leqslant \mathcal{R} _{\mathrm{d},\mathrm{c}}^{\mathrm{s}} = \mathcal{R} _{\mathrm{c},2}^{0}$ and $\mathcal{R} _{\mathrm{s},1}^{\epsilon_1}\leqslant \mathcal{R} _{\mathrm{d},\mathrm{s}}^{\mathrm{s}} = \mathcal{R} _{\mathrm{s},2}^{0}$. When $\mathcal{R} _{\mathrm{c},1}^{\epsilon_1}\in \left[ \mathcal{R} _{\mathrm{d},\mathrm{c}}^{\mathrm{s}}, \mathcal{R} _{\mathrm{d},\mathrm{c}}^{\mathrm{c}}\right]$, there exists an  $\epsilon_2\in\left[0,1\right]$ such that $\mathcal{R} _{\mathrm{c},2}^{\epsilon_2}=\mathcal{R} _{\mathrm{c},1}^{\epsilon_1}$. Based on the monotonicity of $\log _2\left( 1+ax \right) \left( a>0 \right) $ for $x\geqslant 0$, we can deduce that $\epsilon_2<\epsilon_1$, which yields $\mathcal{R} _{\mathrm{s},2}^{\epsilon_2}\geqslant\mathcal{R} _{\mathrm{s},1}^{\epsilon_1}$. The above arguments imply that any rate-tuple on the boundary of $\mathcal{C} _1$ falls within $\mathcal{C} _2$, i.e., $\mathcal{C} _1\subseteq \mathcal{C} _2$. Consequently, we obtain $\mathcal{C} _{\rm{d},\mathrm{f}}\subseteq \mathcal{C} _{\mathrm{1}}\subseteq \mathcal{C} _2\subseteq\mathcal{C} _{\rm{d},\mathrm{i}}$.
\section{Proof of Theorem \ref{uplink_SC_CR}}\label{Proof_uplink_CC_ECR}
\renewcommand{\theequation}{G.\arabic{equation}}
\setcounter{equation}{0}
The uplink sum ECR is written as follows:
\begin{align}\label{G_1}
\mathcal{R} _{\mathrm{u},\mathrm{c},l}^{\mathrm{s}}&=\mathbb{E} \left\{ \sum\nolimits_{m=1}^M{\log _2\left( 1+\gamma _{\mathrm{u},m,l}^{\mathrm{s}} \right) +\log _2\left( 1+\gamma _{\mathrm{u},m^{\prime},l}^{\mathrm{s}} \right)} \right\}  \notag \\
&=\!\sum\nolimits_{m=1}^M\!{\mathbb{E} \left\{ \log _2\!\left( \!1\!+\!\frac{p_{\mathrm{c}}\delta _m}{\left( \mathbf{Q}^{-\mathsf{H}}\mathbf{Q}^{-1} \right) _{m,m}\sigma _{\mathrm{s},l}^{2}} \right)\! \right\}}.
\end{align}
According to \cite[Appendix A]{signalalignment}, the PDF of $\frac{1}{\left( \mathbf{Q}^{-\mathsf{H}}\mathbf{Q}^{-1} \right) _{m,m}}$ is $f_Q\left( x \right)=e^{-x}$, and thus we have
\begin{align}
\mathcal{R} _{\mathrm{u},\mathrm{c},l}^{\mathrm{s}}=\sum\nolimits_{m=1}^M{\int_0^{\infty}{\log _2\left( 1+\frac{p_{\mathrm{c}}\delta _m}{\sigma _{\mathrm{s},l}^{2}}x \right) e^{-x} dx}}.
\end{align}
With the aid of \cite[Eq. (4.337.2)]{integral}, we can derive the result in \eqref{uplink_SC_ECR}. When $p_{\rm{c}}\rightarrow \infty $, we can get \eqref{uplink_SC_ECR_asy} by applying $\lim_{x\rightarrow\infty}\log_2(1+x)\approx\log_2{x}$ and \cite[Eq. (4.331.1)]{integral}.

\section{Proof of Lemma~\ref{uplink_noise_1}}\label{Proof_uplink_noise_1}
\renewcommand{\theequation}{H.\arabic{equation}}
\setcounter{equation}{0}
Let $\bm{\phi}_l\in \mathbbmss{C}^{M\times1}$ and $\mathbf{n}_{{\rm{u}},l}\in \mathbbmss{C}^{M\times1}$ denote the $l$th column of $\bm\Phi$ and $\mathbf{N}_{\rm{u}}$, respectively. Then, we have $\mathbf{n}_{\mathrm{u},l}\sim \mathcal{C} \mathcal{N} \left( 0,\mathbf{I}_M \right) $ and 
\begin{align}
\bm{\phi }_l=&\sum\nolimits_{m=1}^M{\left( \sqrt{\alpha _m\eta _{m}^{-1}p_{\mathrm{c}}}\mathbf{H}_{\mathrm{u},m}\mathbf{w}_{\mathrm{u},m}s_{\mathrm{u},m,l} \right.} \notag\\
&\left.+\sqrt{\alpha _{m^{\prime}}\eta _{m^{\prime}}^{-1}p_{\mathrm{c}}}\mathbf{H}_{\mathrm{u},m'}\mathbf{w}_{\mathrm{u},m'}s_{u,m^{\prime},l} \right)+\mathbf{n}_{\mathrm{u},l}.
\end{align}
As stated before, the communication symbols sent at different time slot are statistically uncorrelated, i.e., $\mathbb{E} \left\{ \mathbf{s}_{\mathrm{u},m}\mathbf{s}_{\mathrm{u},m}^{\mathsf{H}} \right\} =\mathbb{E} \left\{ \mathbf{s}_{\mathrm{u},m^{\prime}}\mathbf{s}_{\mathrm{u},m^{\prime}}^{\mathsf{H}} \right\} =\mathbf{I}_L$. Therefore, we can obtain $\mathbb{E} \left\{\bm{\phi }_l\right\}={\mathbf{0}}$. Based on the principle of signal alignment, we have $\mathbf{H}_{\mathrm{u},m}\mathbf{w}_{\mathrm{u},m}=\mathbf{H}_{\mathrm{u},m'}\mathbf{w}_{\mathrm{u},m'}\triangleq \mathbf{z}_{\mathrm{u},m}$. Furthermore, based on the derivation in \cite{signalalignment} and \cite{Ding_relay}, we have $\mathbf{z}_{\mathrm{u},m}\sim \mathcal{C} \mathcal{N} \left( 0,\mathbf{I}_M \right) $ for $m=1,\ldots,M$ and $\mathbb{E} \left\{\mathbf{z}_{\mathrm{u},m}\mathbf{z}_{\mathrm{u},i}^{\mathsf{H}}\right\}=0$ for $m\ne i$. Taken together, we obtain
\begin{align}
\mathbb{E} \left\{ \bm{\phi} _l\bm{\phi }_{l}^{\mathsf{H}} \right\} &=\left[ \sum\nolimits_{m=1}^M{\left( \alpha _m\eta _{m}^{-1}p_{\mathrm{c}}+\alpha _{m^{\prime}}\eta _{m^{\prime}}^{-1}p_{\mathrm{c}} \right)}+1 \right] \mathbf{I}_M\notag\\
&\triangleq \sigma _{\mathrm{c}}^{2}\mathbf{I}_M,\ l=1,\ldots,L,
\end{align}
and $\mathbb{E} \left\{ \bm{\phi} _l\bm{\phi }_{l'}^{\mathsf{H}} \right\} =0$ for $l\ne l'$. Thus, the aggregate interference-plus-noise $\bm\Phi$ can be regarded as Gaussian noise, with each element having a zero mean and a variance of $\sigma _{\mathrm{c}}^{2}$. On this basis, the expression of SR can be derived by following the steps listed in Appendix~\ref{Appendix:A}. 
\end{appendices}

\bibliographystyle{IEEEtran}
\bibliography{mybib}

\end{document}